\documentclass[preprint,prd,aps,showpacs]{revtex4}
\usepackage{graphicx}
\usepackage{latexsym}
\usepackage{epsfig}
\usepackage{axodraw4j}
\usepackage{color}
\usepackage{amsmath}
\begin{document}
\setcounter{page}{1}
\title{Infra-red Divergences in Light-Front QED and Coherent State Basis}
\author{Jai D. More\footnote{\tt more.physics@gmail.com}}
\author{Anuradha Misra\footnote{\tt misra@physics.mu.ac.in}}
\affiliation{Department of Physics,University of Mumbai,
\\Santa Cruz(E), Mumbai, India-400098}%

\begin{abstract}
We present a next to leading order calculation of electron mass renormalization  in Light-Front Quantum Electrodynamics (LFQED) using old-fashioned time ordered perturbation theory (TOPT). We show that the true infrared divergences  in $\delta m^2$  get canceled up to $O(e^4)$ if  one uses coherent state basis instead of fock basis to calculate the transition matrix elements.
\end{abstract}
\pacs{11.10.Ef,12.20.Ds,12.38.Bx }
\date{\today}
\maketitle

\section{INTRODUCTION}
It is well known that in Quantum Electrodynamics (QED) the infra-red (IR) divergences get cancelled in suitably defined cross sections by virtue of the famous Bloch-Nordseick theorem\cite {BLOCH37}. According to this theorem, the divergences in virtual processes get canceled when the contribution of real photon emission is added. It is to be noted that this cancellation takes place at the level of cross-sections and not at the level of amplitudes. The divergences at the amplitude level arise due to inappropriate choice of initial and final states. In LSZ formulation, the dynamics of incoming and outgoing particles in a scattering event is described by the free Hamiltonian and therefore, the initial and final states used to calculate the transition matrix elements are taken to be the Fock states. However, in an actual experiment, due to the finite size of the detector, the charged particle can be accompanied by any number of photons. The Bloch-Nordseick mechanism takes into account all states with any number of soft photons below experimental resolution thus leading to cancellation of divergences. The issue of cancellation of IR divergences at the amplitude level was addressed by Chung\cite{CHU65} who showed that the divergences in matrix elements are eliminated to all orders in perturbation theory if one chooses the initial and final states to be charged particles with a suitable superposition of an infinite number of photons. Kulish and Faddeev\cite{KUL70} defined the asymptotic states by means of an asymptotic Hamiltonian. They were the first to show that in QED, the asymptotic Hamiltonian does not coincide with the free Hamiltonian. Kulish and Fadeev(KF) constructed the asymptotic Hamiltonian $V_{as} $ for QED thus modifying  the asymptotic condition to introduce a new space of asymptotic states given by\begin{equation} 
\vert n;\pm \rangle = \Omega^A_{\pm}\vert n\rangle
\end{equation}
where $\Omega^A_{\pm} $ is the asymptotic evolution operator and $\vert n\rangle$ is the Fock state. $\Omega^A_{\pm} $ is defined by 
\begin{eqnarray}\label{omega}
\Omega^A_{\pm} = T~exp\bigg[-i\int_{\mp}^0 V_{as}(t) dt \bigg]
\end{eqnarray}

KF further modified the definition of S matrix and showed that it is free of IR divergences. In a nutshell, the method of asymptotic dynamics proposed by Kulish and Fadeev replaces the free Hamiltonian by an asymptotic Hamiltonian which takes into account the long range interaction between incoming and outgoing states and can be used to construct a set of coherent states as the asymptotic states. The transition matrix elements formed by using these states are then free of infra-red divergences.

The KF method was used by Nelson and Butler\cite{BUT78, NEL81, NEL181} to generate a set of asymptotic states in the asymptotic region of perturbative Quantum Chromodynamics(pQCD). The asymptotic states constructed were shown to lead to cancellation of IR divergences in certain matrix elements in lowest order in pQCD.  Greco {\it etal}\cite{GRECO78} constructed a coherent state approach for non-abelian gauge theories and showed that matrix elements between coherent states of definite color are finite and factorized in the fixed angle regime. KF method was applied to QCD by Dahmein and Steiner\cite{DAH81} also who showed that the leading logarithmic behavior of the mass shell form factor can be derived from asymptotic quark gluon part of the QCD Hamiltonian.

Relevance of coherent state formalism in light-front field theory (LFFT) was first discussed by Harindranath and Vary\cite{HARI88} who showed that a coherent state may be a valid vacuum in LFFT. Later, it was shown \cite{VARY99} in the context of LF Schwinger Model that the physical vacuum is gauge invariant superposition of coherent states 
of dynamical gauge field zero mode.

A coherent state formalism for LFQED was developed by one of us in Ref.\cite{ANU94}, henceforth referred to as $\mathcal I$, as a possible method to deal with the {\it true} IR divergences of LFFT. These {\it true} IR divergences are the bona-fide divergences of equal-time field theory and appear when both $k^+$ and ${\bf k_\perp}$ approach zero. In addition to these, there are additional IR divergences in LFFT called the {\it spurious} IR divergences as they are just a manifestation of ultra-violet divergences of equal-time theory.  It was shown in $\mathcal I$ that the true IR divergences in one loop vertex correction are eliminated when the transition matrix element is calculated between the coherent states in place of Fock states. Subsequently, it was proposed\cite{ANU96} to use the coherent state basis constructed in $\mathcal I$ for calculation of Hamiltonian matrix elements in Discrete Light Cone Quantization(DLCQ) method of bound state calculation as a possible way to avoid the vanishing energy denominators and the resulting true IR divergences. The method was applied to obtain the light-cone Schrodinger equation for positronium using coherent state basis and to demonstrate the absence of Coulomb singularity therein. The method of asymptotic dynamics has also been applied to LFQCD\cite{ANU00} to obtain a set of coherent states and it has been shown to lead to cancellation of IR divergences appearing due to vanishing energy denominators (which are actually the true IR divergences) in qqg vertex correction at one loop level.

KF method leads to cancellation of IR divergences in QED to all orders. However, it is well known that the Bloch-Nordseick theorem does not hold in QCD and therefore, in this case one does not expect to construct an all order proof of cancellation of IR divergences along the lines of KF method. Basically, the non-cancellation of IR divergences in QCD stems from the fact that asymptotic states here are bound states of quarks and anti-quarks and therefore the asymptotic Hamiltonian to be used in KF method should contain the confining potential and is not just the asymptotic Hamiltonian of QCD. An "improved" method of asymptotic dynamics was introduced by McMullan {\it etal} \cite{HOR98, HOR99, HOR00} which takes into account the separation of particles also. The improved method has also been discussed in the context of LFQED and LFQCD\cite{ANU05}.

In this work, we calculate fermion self energy in LFQED up to $O (e^4) $. We extend the analysis of $\mathcal I$ to include the instantaneous interaction also in the construction of asymptotic Hamiltonian and the resulting coherent state basis. We show that the true IR divergences in electron mass renormalization are cancelled up to $O (e^4) $ if one uses coherent state basis for evaluating the transition matrix elements. 

Conventionally, LF quantization is performed in light-front gauge $A^+=0$ due to its many advantages when applied to non-abelian theory \cite{PREM01}  - specifically  due to the absence of ghost fields. In this work on QED, we have also used light-front gauge.  There has been some discussion addressing the question of gauge independence of LFQED calculations  in literature \cite{HILLER04}. In a recent work, gauge independence of non perturbative calculation of electron's anomalous magnetic moment  has been verified \cite{HILLER11}. However, in this work we have not addressed the issue of gauge dependence of our results as we plan to do this  in a future work. 

The cancellation of IR divergences in covariant QED has been established to all orders both in the amplitude as well as the cross-section approach. Various authors \cite{ANUS05, BAKKER95, SCHO98, BAKKER05} have addressed the issue of equivalence of covariant and Light-Front formalism of field theory and therefore it may be considered unnecessary to address the cancellation of IR divergence in LFFT. However, the divergence structure of LFFT's is different from that of covariant formalism and there are issues present that still need to be addressed \cite{PERRY97}.  In particular, it is important to differentiate between true and spurious IR divergences. As discussed in Ref. \cite{ANU94, ANU96, ANU00, ANU05} a coherent state approach in LFFT is interesting due to following reason. LFFT's being based on a Hamiltonian approach, a coherent state method in LFFT may be useful from the point of view of extracting information about the artificial confining potential which is needed in LF bound state calculations\cite{WILSON94}. It is well known that IR divergences do not cancel in QCD and the reason within the coherent state formalism is that the asymptotic states are not the asymptotic states of QCD but are bound states of quarks and antiquarks. In other words, if we use appropriate Hamiltonian of bound states as the asymptotic Hamiltonian and develop a coherent state approach based on it, then this approach would lead to cancellation of IR divergences in QCD as well. It will be worthwhile to understand this connection between cancellation/non-cancellation of IR divergences and the form of asymptotic Hamiltonian. The hope is that by understanding the structure of IR divergences we may be able to get some insight into the form of artificial confining potential mentioned in Ref. \cite{WILSON94} which can then be used to perform the bound state calculations. 

The plan of the paper is as follows: In Section II, we present the Hamiltonian of LFQED and calculate the $O (e^2)$ electron mass renormalization using light-cone -time-ordered perturbation theory(LCTOPT) in the standard Fock basis. We demonstrate the appearance of true IR divergences in the form of vanishing light-cone energy denominators. In Section III, we obtain the form of coherent states using the method of asymptotic dynamics. In Section IV, we calculate $\delta m^2$ in lowest order using the coherent state basis and show that the extra contributions due to emission and absorption of soft photons indeed cancel the IR divergences in $\delta m^2$. In Section V, we calculate $\delta m^2$ up to $O(e^4)$ in Fock basis and identify the IR divergences in it. In Section VI we perform the same calculation in coherent state basis and show the cancellation of IR divergences in this basis. Section VII contains a summary and discussion of our results. In Appendix A, we set the notations and conventions and give some useful relations. Appendix B contains some useful properties of coherent states. Appendix C and D contain the details of the calculation of transition matrix element in Fock basis and coherent state basis respectively.

\section{PRELIMINARIES}
\subsection{Light-Front QED Hamiltonian}
The light-front QED Hamiltonian in the light-front gauge $(A^+=0)$ expressed in terms of independent degrees of freedom is given by  \cite {MUS91,BRODSKY98}
\begin{equation}
P^-= H \equiv H_0 + V_1 + V_2 + V_3 \;,
\end{equation}
where
\begin{equation}
H_0= \int d^2 {\bf x}_\perp dx^- \{\frac{i}{2} \bar{\xi}\gamma^-
\stackrel{\leftrightarrow}{\partial}_-\xi + \frac{1}{ 2} (F_{12})^2-\frac{1}{2}a_+\partial_- \partial_k a_k \}
\end{equation}
is the free Hamiltonian,
\begin{equation}
V_1=e \int d^2{\bf x}_\perp dx^- \bar{\xi} \gamma^{\mu}\xi a_\mu\;
\end{equation}
is the standard $O(e)$ three point interaction,
\begin{eqnarray}
V_2=-\frac{i}{4}e^2\int d^2{\bf x}_\perp dx^-dy^-\epsilon(x^--y^-)(\bar \xi a_k \gamma^k)(x)\gamma^+(a_j\gamma^j\xi)(y)
\end{eqnarray}
is an $O(e^2)$ non-local effective four-point interaction corresponding to an instantaneous 
fermion exchange and
\begin{eqnarray}
V_3=-\frac{e^2}{4}\int d^2{\bf x}_\perp dx^-dy^-(\bar\xi \gamma^+\xi)(x) \vert x^--y^-\vert (\bar \xi\gamma^+\xi)(y)
\end{eqnarray}
is an $O(e^2)$ non-local effective four-point interaction corresponding to an instantaneous photon exchange. $V_2$ and $V_3$ are drawn as four-point interactions and a hash mark  is drawn on the line representing the instantaneous particle. $\xi(x)$ and $a_\mu(x)$ can be expanded in terms of creation and annihilation operators as
\begin{align}
\xi(x)=\int\frac{d^2 {\bf p}_\perp}{(2\pi)^{3/2}}\int 
\frac{dp^+}{\sqrt{2p^+}}\sum_{s=\pm\frac{1}{2}}[u(p,s)e^{-i(p^+x^--\bf{p_\perp x_\perp)}}b(p,s,x^+)
                                             \nonumber\\
+v(p,s)e^{i(p^+x^--{\bf p}_\perp x_\perp)}d^\dagger(p,s,x^+)],
\end{align}
\begin{align}
a_\mu(x)=\int\frac{d^2{\bf q}_\perp}{(2\pi)^{3/2}}\int
\frac{dq^+}{\sqrt{2q^+}}\sum_{\lambda=1,2}\epsilon^\lambda_\mu(q)
[e^{-i(q^+x^--{\bf q}_\perp x_\perp)}a(q,\lambda,x^+)
+e^{i(q^+x^--{\bf q}_\perp x_\perp)}a^\dagger(q,\lambda,x^+)],
\end{align}
where
\begin{equation}\label{anticommutation}
\{b(p,s),b^\dagger(p^\prime,s^\prime)\}=\delta(p^+-p^{\prime+})\delta^2({\bf 
p_\perp-p^\prime_\perp})\delta_{ss^\prime}
=\{d(p,s),d^\dagger(p^\prime,s^\prime)\},
\end{equation}
\begin{equation}\label{commutation}
[a(q,\lambda),a^\dagger(q^\prime,\lambda^\prime)]=\delta(q^+-q^{\prime+})\delta^2({\bf q_\perp-q^\prime_\perp}) \delta_{\lambda\lambda^\prime}.
\end{equation}
These relations hold at equal light-front time $x^+$.
In terms of these momentum-space operators, the free Hamiltonian has the form
\begin{align}\label{free Ham}
H_0=\int d^2{\bf p}_\perp dp^+\bigg[\frac{{\bf p}_\perp^2 + m^2}{2p^+}\sum_{s=\pm \frac{1}{2}} (b^\dagger(p,s)b(p,s)+d^\dagger(p,s)d(p,s))
+\frac{{\bf p}_\perp^2}{2p^+}\sum_{\lambda=1,2} a^\dagger(p,s)a(p,s)\bigg]
\end{align}
Similarly, $V_1$  has the form 
\begin{align}\label{V_1}
V_1=e\int d^2 {\bf x}_\perp dx^-\int[dp][d\overline{p}][dk]\sum_{s,s^\prime,\lambda}[e^{i\overline{p}\cdot x}\overline{u}(\overline{p},s^\prime)b^\dagger(\overline{p},s^\prime)+e^{-i\overline{p}\cdot x}\overline{v}(\overline{p},s^\prime)d(\overline{p},s^\prime)]\nonumber\\
\times\gamma^\mu[e^{-ip\cdot x} u(p,s) b(p,s)+e^{ip\cdot x} v(p,s)d^\dagger(p,s)] \epsilon^\lambda_\mu(k)[e^{-ik\cdot x}a(k,\lambda)+e^{ik\cdot x}a^\dagger(k,\lambda)],
\end{align}
where
\begin{equation}
\int [dp] \equiv \int_{-\infty}^{\infty} {d^2{\bf p}_\perp \over {(2
\pi)^{3\over 2}}} \int_0^\infty{dp^+ \over {\sqrt{2p^+}}}
\end{equation}
$V_2$ and $V_3$ are given by following expressions:
\begin{align}\label{V_2}
V_2=&-\frac{ie^2}{4}\int d^2{\bf x}_\perp dx^-dy^- dl[dp][d\overline{p}][dk][d\overline{k}]\sum_{s,s^\prime,\lambda,\lambda^\prime}\bigg[-\frac{i}{\pi l}\bigg]e^{il(x^--y^-)}\nonumber\\
&[e^{i p\cdot x}\overline{u}(p,s)b^{\dagger}(p,s)+e^{-ip\cdot x}\overline{v}(p,s)d(p,s)]{\epsilon\llap/}^\lambda(k)[e^{-ik\cdot x}a(k,\lambda)+e^{ik\cdot x}a^\dagger(k,\lambda)]\gamma^+\nonumber\\
& {\epsilon\llap/}^{\lambda^\prime}(\overline{k})[e^{-i\overline{p}\cdot y}u(\overline{p},s^\prime)b(\overline{p},s^\prime)+e^{i\overline{p}\cdot y}v(\overline{p},s^\prime)d^\dagger(\overline{p},s^\prime)]
[e^{-i\overline{k}\cdot y}a(\overline{k},\lambda^\prime)+e^{i\overline{k}\cdot y}a^\dagger(\overline{k},\lambda^\prime)],
\end{align}
\begin{align}\label{V_3}
V_3=&\frac{e^2}{4}\int d^2{\bf x}_\perp dx^- dy^- dl[dp][d\overline{p}][dk][d\overline{k}]\sum_{s,s^\prime,\sigma,\sigma^\prime}\frac{e^{il(x^--y^-)}}{\pi l^2}[e^{i \overline p\cdot x}\overline{u}(\overline p,s^\prime)b^{\dagger}(\overline p,s^\prime)\;\nonumber\\
&+e^{-i\overline p\cdot x}\overline{v}(\overline p,s^\prime)d(\overline p,s^\prime)]\gamma^+[e^{-i{p}\cdot x}u(p,s)b(p,s)+e^{i p\cdot x}v(p,s)d^\dagger(p,s)][e^{i\overline k\cdot y}\overline u(\overline k,\sigma^\prime) b^\dagger (\overline k,\sigma^\prime)
\nonumber\\
&+e^{-i\overline k\cdot y}\overline v(\overline k,\sigma^\prime) d(\overline k,\sigma^\prime)]\gamma^+[e^{-ik\cdot y}u(k,\sigma)b(k,\sigma)+e^{ik\cdot y}v(k,\sigma)d^\dagger(k,\sigma)]
\end{align}
where $y=(x^+,y^-,{\bf x}_\perp)$.
\subsection{Electron Mass Renormalization in Light-Front QED}
In light-front time ordered perturbation theory, the transition matrix is given by the perturbative expansion
\begin{equation}
T= V + V {1 \over {p^--H_0}}V + \cdots
\end{equation}
The electron mass shift is obtained by calculating $T_{pp}$ which is the matrix element of the above series between the initial and the final electron states $\vert p,s \rangle$ and $\vert p,\sigma \rangle$ and it is given by \cite {MUS91}, 
\begin{align}\label{deltam}
\delta m^2= p^+ \sum_{s} T_{pp}
\end{align}
Note that only $\sigma = s$ contributes, as the fermion self energy is diagonal in spin.

We expand $T_{pp}$ in powers of $e^2$ as  
\begin{align}
T_{pp}=T^{(1)}+T^{(2)}+\cdots
\end{align}
In general, $T^{(n)}$ gives the $O(e^{2n})$ contribution to electron self energy correction. Here, the initial (or final) electron momentum is
\begin{align}
p=\biggl[p^+,\frac{{\bf p}_\perp^2+m^2}{2p^+},{\bf p}_\perp\biggr].
\end{align}   
In particular, $O(e^2)$ correction is obtained from
\begin{align}\label{def diag1}
T^{(1)}_{pp}\equiv T^{(1)}( p,p)=\langle p,s \vert V_1 \frac{1} {p^- - H_0}V_1\vert p,s \rangle+\langle p,s \vert V_2\vert p,s \rangle\nonumber
\end{align}
\begin{figure}[h]
\includegraphics[scale=0.6]{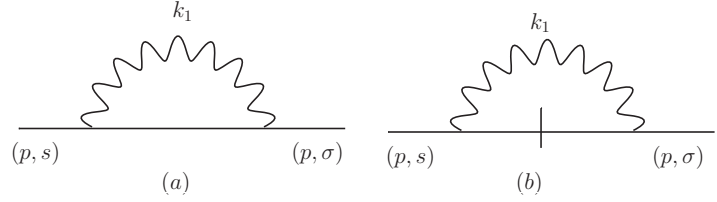}
\label{fig1}
\caption{Diagrams for $O(e^2)$ self energy correction in fock basis corresponding to $T_1$}
\end{figure} \\  
Note that
\begin{align}
T^{(1)}_{pp} \equiv T^{(1)}(p,p)=T_{1a}+T_{1b}\;
\end{align}
where $T_{1a}$ and $T_{1b}$ are $O(e^2)$ contributions from standard three point vertex and the 4-point instantaneous vertex and are represented by the diagrams in Fig. 1(a) and Fig. 1(b) respectively. We are interested in the true IR divergences which arise due to vanishing energy denominators in TOPT \cite{ANU94}. It is obvious that $T_{1b}$ cannot have such IR divergences as there are no energy denominators involved and hence this term is not required in  our discussion. Neglecting $T_{1b}$, $O(e^2)$ transition matrix element contributing to fermion self energy reduces to
\begin{eqnarray}
T_{1a}(p,p)=\langle p,s \vert V_1 \frac{1} {p^- - H_0}V_1\vert p,s \rangle
\end{eqnarray} 
To calculate $T_{1a}$ we insert  two complete sets of states so that the above equation becomes
\begin{eqnarray}
T_{1a}(p,p)=\sum_{spins}\int \prod_{i=1}^2 d^3p_i^\prime  d^3k_i^\prime \langle p,s \vert V_1 \vert p_1^\prime,s_1^\prime,k_1^\prime,\lambda_1^\prime\rangle\langle p_1^\prime,s_1^\prime,k_1^\prime,\lambda_1^\prime\vert\frac{1} {p^- - H_0}\vert p_2^\prime,s_2^\prime,k_2^\prime,\lambda_2^\prime\rangle \nonumber\\\langle p_2^\prime,s_2^\prime, k_2^\prime,\lambda_2^\prime\vert V_1\vert p,s \rangle
\end{eqnarray}
Substituting for $V_1$ from Eq.~(\ref{V_1}) and using Eqs.~(\ref{deltam}) and (\ref{uubar}) we obtain
\begin{align}
\delta m^2_{1a}=\frac{e^2}{2(2 \pi)^3}\int{d^2{\bf k}_{1\perp}}\int\frac{dk_1^+}{k_1^+p_1^+} \frac{Tr[\epsilon\llap/^\lambda(k_1) (\not p_1+m) \epsilon\llap /^\lambda(k_1) (\not p+m)]}{4(p^--p_1^--k_1^-)}
\end{align}
where \quad  $p_1=p-k_1$.\\
Calculating the trace,  using Eq.~(\ref{relation1}) for energy denominator and taking the limit $k_1^+ \rightarrow 0$, ${\bf k}_{1\perp} \rightarrow 0$,  we finally obtain
\begin{eqnarray}\label{diag1}
{(\delta m^2_{1a})}^{IR}=-\frac{e^2}{(2\pi)^3}\int{d^2{\bf k}_{1\perp}}\int\frac{dk_1^+}{k_1^+}\frac{(p\cdot \epsilon(k_1))^2}{(p\cdot k_1)}
\end{eqnarray}
Note that the denominator vanishes as  $k_1^+ \rightarrow 0$, ${\bf k}_{1\perp} \rightarrow 0$ leading to true IR divergences \cite{ANU94}. 
\vskip 0.25cm
\section{INFRARED DIVERGENCES AND THE COHERENT STATE BASIS}

It was shown in $\mathcal I$, that the true IR divergences in one loop vertex correction get cancelled if one uses coherent state basis in LFQED. We will prove the same result for electron mass renormalization in  Section IV. For that purpose, we will now obtain the form of coherent states in light-front formalism by the method used in Ref.\cite{KUL70} for equal time theory. In $\mathcal I$, only three point vertex was used to obtain the asymptotic Hamiltonian and the corresponding coherent state basis. We extend the formalism developed in $\mathcal I$ by obtaining asymptotic limit of four point instantaneous interaction also. \\
\indent The light-front time dependence of the interaction Hamiltonian is given by
\begin{displaymath}
H_I(x^+)=V_1(x^+)+V_2(x^+)+V_3(x^+)
\end{displaymath}
where \cite{ANU94}
\begin{align}
V_1(x^+)= e\sum_{i=1}^4 \int d\nu_i^{(1)}[ e^{-i \nu_i^{(1)} x^+} {\tilde h}_i^{(1)}(\nu_i^{(1)})
+ e^{i \nu_i^{(1)} x^+}
{\tilde h}^{(1)\dagger}_i (\nu_i^{(1)})]
\end{align}
 ${\tilde h}^{(1)}_i(\nu^{(1)}_i)$ are three point QED interaction vertices :
\begin{equation}
{\tilde h}^{(1)}_1 = \sum_{s,s^{\prime},\lambda} b^{\dagger}(\overline p,s^\prime)b(p,s)
a(k,\lambda) \overline u(\overline
p,s^\prime)\gamma^{\mu}u(p,s)\epsilon^{\lambda}_{\mu}\;,
\end{equation}
\begin{equation}
{\tilde h}^{(1)}_2 = \sum_{s,s^{\prime},\lambda} b^{\dagger}(\overline p,s^\prime)
d^{\dagger}(p,s)
a(k,\lambda) \overline u(\overline
p,s^\prime)\gamma^{\mu}v(p,s)\epsilon^{\lambda}_{\mu}\;,
\end{equation}
\begin{equation}
{\tilde h}^{(1)}_3 = \sum_{s,s^{\prime},\lambda} d(\overline p,s^\prime)b(p,s)
a(k,\lambda) \overline v(\overline
p,s^\prime)\gamma^{\mu}u(p,s)\epsilon^{\lambda}_{\mu}\;,
\end{equation}
\begin{equation}
{\tilde h}^{(1)}_4 = \sum_{s,s^{\prime},\lambda} d^{\dagger}(\overline p,s^\prime)d(p,s)
a(k,\lambda) \overline v(\overline
p,s^\prime)\gamma^{\mu}v(p,s)\epsilon^{\lambda}_{\mu}\;,
\end{equation}
and
$\nu_i^{(1)}$ is  the light-front energy transferred at the vertex ${\tilde h}_i^{(1)}$.  For example,
\begin{equation}
\nu_1^{(1)} = p^- + k^- - \overline p^- = {p \cdot k \over p^++k^+}
\end{equation}
is the energy transfer at $ee\gamma$ vertex. The integration measure is given by
\begin{equation}
\int d\nu^{(1)}={1\over{{(2\pi)}^{3/2}}}\int{{[dp][dk]}\over{\sqrt{2\overline p^+}}} \;,
\end{equation}
$\overline p^+$ and $\overline {\bf p}_{\perp}$ being fixed at each vertex by momentum
conservation.\\
\indent At asymptotic limits, non-zero contributions to $V_1(x^+)$ come from regions
where $\nu_i^{(1)} \rightarrow 0$. It is easy to see that $\nu_2^{(1)}$ and $\nu_3^{(1)}$ are always non-zero and therefore, $\tilde h_2$ and $\tilde h_3$ do not appear in the asymptotic Hamiltonian. Thus, the 3-point asymptotic Hamiltonian is defined by the following expression \cite {ANU94}
\begin{eqnarray}\label{V_1as}
V_{1as}(x^+) = e \sum_{i=1,4} \int d\nu_i^{(1)} \Theta_\Delta(k)
 [ e^{-i \nu_i^{(1)} x^+} \tilde h_i^{(1)}(\nu_i^{(1)})
+ e^{i \nu_i^{(1)} x^+}
\tilde h^ {\dagger}_i (\nu_i^{(1)})] \;
\end{eqnarray}
where $\Theta_{\Delta}(k)$ is a function which takes value 1 in the asymptotic region and is zero elsewhere.

As shown in $\mathcal I$, for $V_1$ we can define the asymptotic region to consist of all points in the phase space for which
\begin{equation}
{{p \cdot k} \over p^+} < \Delta E     \;,
\end{equation}
where $\Delta E$ is an energy cutoff which may be chosen to be the experimental resolution.  For simplicity, we shall choose a frame ${\bf p}_\perp = 0 $. In this frame, the above condition reduces to
\begin{equation}\label{asyrel}
{{p^+{\bf k}_\perp^2} \over 2k^+} + {m^2k^+ \over 2p^+} < \Delta \;,
\end{equation}
where $\Delta = p^+\Delta E$.

Thus, for all the points satisfying Eq.~(\ref{asyrel}), $\nu_1^{(1)}$ and $\nu_4^{(1)}$ can be approximated by zero. This implies that in  this region, the  asymptotic Hamiltonian is different from the free Hamiltonian. For the present purpose, i.e. in order to eliminate the true IR divergences, we find it sufficient to choose a subregion of the above mentioned region as the asymptotic region. We define this subregion to be consisting of all points $(k^+, {\bf k}_\perp)$ satisfying:
\begin{equation}
{\bf k}_\perp ^2 < {{k^+ \Delta} \over{p^+}}  \;,
\end{equation}
\begin{equation}
k^+ < {{p^+ \Delta} \over {m^2}}   \;.
\end{equation}
This choice of the asymptotic region leads to the asymptotic interaction Hamiltonian defined by Eq.~(\ref{V_1as}) with
\begin{equation}\label{theta1}
 \Theta_\Delta(k)=\theta\bigg({{k^+\Delta} \over p^+} - {\bf k}_\perp^2\bigg)
\theta\bigg({{p^+\Delta} \over m^2} - k^+\bigg)
\end{equation}
\indent Contribution to asymptotic Hamiltonian from the four point instantaneous interaction can be obtained by taking $\vert x^+ \vert \rightarrow  \infty$ limit in $ V_2(x^+)$. $ V_2(x^+)$ is given by 
\begin{eqnarray}\label{V_2(X^+)}
V_2(x^+)= e^2 \sum_{i=1}^8 \int d\nu_i^{(2)}[ e^{-i \nu_i^{(2)} x^+} {\tilde h}_i^{(2)}(\nu_i^{(2)})+ e^{i \nu_i^{(2)} x^+}
{\tilde h}^{(2)\dagger}_i (\nu_i^{(2)})]{1\over{2 (\pm p^+\pm k_1^+)}}
\end{eqnarray}
where ${\tilde h}^{(2)}_i(\nu^{(2)}_i)$ are 4-point instantaneous fermion exchange vertices. For example,
\begin{eqnarray}
{\tilde h}^{(2)}_1 = \sum_{s,s^{\prime},\lambda_1,\lambda_2} b^{\dagger}(\overline p,s^\prime)b(p,s)a(k_1,\lambda_1)a(k_2,\lambda_2) \overline u(\overline p,s^\prime)\gamma^{\mu}\gamma^+ \gamma^{\nu}u(p,s)\epsilon^{\lambda_1}_{\mu}(k_1) \epsilon^{\lambda_2}_{\nu}(k_2)\;,\nonumber
\end{eqnarray}
One can write the remaining seven terms in a similar manner. $\nu_i^{(2)}$ is  the light front energy transferred at the vertex ${\tilde h}_i^{(2)}$. For example, in Eq.~(\ref{V_2(X^+)}) 
\begin{equation}\label{asreln2}
\nu_2^{(2)} = p^- - k_1^-+ k_2^- - \overline p^- = -\frac{p \cdot k_1-p \cdot k_2+k_1 \cdot k_2}{ p^+-k_1^++k_2^+}
\end{equation}
and the integration measure is 
\begin{equation}
\int d\nu^{(2)} =
\frac{1}{{(2 \pi)}^{3/2}} \int\frac{[dp][dk_1][dk_2]}{\sqrt{2\overline p^+}} \;.
\end{equation} 
\indent At asymptotic limits, non-zero contributions to $V_2(x^+)$ come from regions where $\nu_i^{(2)} \rightarrow 0$. It can be  shown easily that  $\nu_2^{(2)}$ and $\nu_8^{(2)}$ vanish when $k_1^+= k_2^+$ and  ${\bf k}_{1\perp}={\bf  k}_{2\perp}$, while the rest of the six $\nu_i^{(2)}$'s are always non-zero. Thus, the asymptotic Hamiltonian  for $V_2$ is defined by the following expression
\begin{eqnarray}\label{V_2as}
V_{2as}(x^+)=e^2 \sum_{i=2,8} \int d\nu_i^{(2)}\delta^3(k_1-k_2)[ e^{-i \nu_i^{(2)} x^+} {\tilde h}_i^{(2)}(\nu_i^{(2)})+ e^{i \nu_i^{(2)} x^+}
{\tilde h}^{(2)\dagger}_i (\nu_i^{(2)})]{1\over{2 (p^+-k_1^+)}}
\end{eqnarray}
Similarly,  $V_{3as}(x^+)$ is obtained by taking the limit $\vert x^+ \vert \rightarrow  \infty$ in $V_3(x^+)$, where
\begin{eqnarray}
V_3(x^+)= e^2 \sum_{i=1}^{8} \int d\nu_i^{(3)}[ e^{-i \nu_i^{(3)} x^+} {\tilde h}_i^{(3)}(\nu_i^{(3)})+ e^{i \nu_i^{(3)} x^+}
{\tilde h}^{(3)\dagger}_i (\nu_i^{(3)})]{1\over{2 (\pm p^+\pm\overline p^+)^2}}
\end{eqnarray}
Here, ${\tilde h}^{(3)}_i(\nu^{(3)}_i)$ are 4-point instantaneous photon exchange vertices. 
One can easily verify that $\nu_i^{(3)}$'s are always non-zero for all i. Hence,  $V_3(x^+)$ is zero in the asymptotic limit and does not contribute to the asymptotic Hamiltonian. \\
\indent The asymptotic states can be defined in the usual manner by 
\begin{equation}
\vert n \colon coh \rangle = \Omega_{\pm}^A \vert{n}\rangle    \;,
\end{equation}
where $\vert{n}\rangle$ is a Fock state and $\Omega_{\pm}^A$ are the asymptotic M\"oller operators defined by
\begin{equation}
\Omega_{\pm}^A = T~exp\biggl[ -i \int^0_{\mp}[ V_{1as}(x^+)+ V_{2as}(x^+)]dx^+ \biggr] \;.
\end{equation}
Carrying out the standard procedure\cite{KUL70} of substituting $k^+=0$,
${\bf k}_{\perp}=0$ in all
the slowly varying functions of $k$, and carrying out the $x^+$
integration, we arrive at the following expression for the asymptotic states:
\begin{align}
\Omega_{\pm}^A \vert n\colon p_i \rangle=&exp\biggl[-e\int{dp^+d^2{\bf p}_\perp}\int\sum_{\lambda=1,2}
\frac{d^2{\bf k}_\perp}{(2\pi)^{3/2}}\int{{dk^+}\over{\sqrt{2k^+}}}\nonumber\\&[f(k,\lambda:p) a^\dagger(k,\lambda)- f^*(k,\lambda:p)a(k,\lambda)]\nonumber\\ &+e^2\int{dp^+d^2{\bf p}_\perp}\int \sum_{\lambda_1=1,2}\frac{d^2{\bf k} _{1\perp}}{(2\pi)^{3/2}}\int{{dk_1^+}\over{\sqrt{2k_1^+}}}\int\sum_{\lambda_2=1,2}{{d^2{\bf k}_{2\perp}} \over{(2\pi)^{3/2}}}\int{{dk_2^+}\over{\sqrt{2k_2^+}}}\nonumber
\end{align}
\begin{align}
[g_1(k_1,k_2,\lambda_1,\lambda_2 \colon p) a^\dagger(k_2,\lambda_2)a(k_1,\lambda_1)-g_2(k_1,k_2,\lambda_1,\lambda_2 \colon p)a(k_2,\lambda_2)a^\dagger(k_1,\lambda_1)]\rho(p)\biggr]\vert n \colon p_i \rangle \;
\end{align}
where
\begin{equation}
f(k,\lambda \colon p) = {{p_\mu\epsilon_\lambda^\mu(k)} \over {p\cdot k}}
\theta\bigg(\frac{k^+\Delta}{p^+}-{\bf k}_\perp^2\bigg)
\theta\bigg(\frac{p^+\Delta}{m^2}-k^+\bigg) \;,
\label{eq:theta2}
\end{equation}
\begin{equation}
f(k,\lambda \colon p) = f^*(k, \lambda \colon p) \;,
\end{equation}
\begin{align}
g_1(k_1,k_2,\lambda_1,\lambda_2 \colon p)=-\frac{4p^+}{p \cdot k_1-p \cdot k_2+k_1 \cdot k_2} \delta^{3}(k_1-k_2)\nonumber\\
g_2(k_1,k_2,\lambda_1,\lambda_2 \colon p)=\frac{4p^+}{p \cdot k_1-p \cdot k_2-k_1 \cdot k_2}\delta^{3}(k_1-k_2)
\end{align}
\begin{equation}
\rho(p) = \sum_n\bigl[b_n^\dagger(p)b_n(p) - d_n^\dagger(p)d_n(p) \bigr]\;.
\end{equation}
Applying the operator $\rho(p)$ on  the Fock state, we finally obtain
\begin{align}
\Omega_{\pm}^A \vert n \colon p_i \rangle=
&exp\biggl[-e\int \sum_{\lambda=1,2}
\frac{d^2{\bf k}_\perp}{(2\pi)^{3/2}}\int \frac{dk^+}{\sqrt{2k^+}}[f(k,\lambda,p) a^\dagger(k,\lambda) - f^*(k,\lambda,p)a(k,\lambda)]\nonumber\\
&+ e^2\int \sum_{\lambda_1=1,2} \frac{d^2{\bf k}_{1\perp}}{(2\pi)^{3/2}}\int\frac{dk_1^+}{\sqrt{2k_1^+}}\int \sum_{\lambda_2=1,2}\frac{d^2{\bf k}_{2\perp}}{(2\pi)^{3/2}}\int\frac{dk_2^+}{\sqrt{2k_2^+}}\nonumber
\end{align}
\begin{align}
[g_1(k_1,k_2,\lambda_1,\lambda_2 \colon p) a^\dagger(k_2,\lambda_2)a(k_1,\lambda_1) - g_2(k_1,k_2,\lambda_1,\lambda_2 \colon p)a(k_2,\lambda_2)a^\dagger(k_1,\lambda_1)\biggr]
\vert n \colon p_i \rangle \;
\end{align}
In particular, the one fermion coherent state  is given by
\begin{align}
\vert p,\sigma \colon f(p) \rangle=&exp\biggl[-e\int \sum_{\lambda=1,2}
 \frac{d^2{\bf k}_\perp}{(2\pi)^{3 \over 2}}\int \frac{dk^+}{\sqrt{2k^+}}[f(k,\lambda \colon p) a^\dagger(k,\lambda) - f^*(k,\lambda \colon p)a(k,\lambda)]\nonumber\\
&+ e^2\int \sum_{\lambda_1=1,2} {{d^2{\bf k}_{1\perp}} \over {(2\pi)^{3/2}}}\int{{dk_1^+}\over{\sqrt{2k_1^+}}}
\int \sum_{\lambda_2=1,2}{{d^2{\bf k}_{2\perp}} \over {(2\pi)^{3/2}}}\int{{dk_2^+}\over{\sqrt{2k_2^+}}}\nonumber
\end{align}
\begin{align}\label{state1}
[g_1(k_1,k_2,\lambda_1,\lambda_2 \colon p) a^\dagger(k_2,\lambda_2)a(k_1,\lambda_1) - g_2(k_1,k_2,\lambda_1,\lambda_2 \colon p)a(k_2,\lambda_2)a^\dagger(k_1,\lambda_1)\biggr]
\vert p,\sigma \rangle \;
\end{align}
Some useful properties of these coherent states are listed in Appendix B. 
\vskip 0.5cm
\section{ELECTRON MASS RENORMALIZATION upto $O(e^2)$ IN COHERENT STATE BASIS}
\begin{figure}[h]
\includegraphics[scale=0.6]{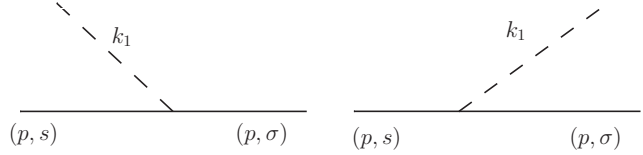}
\label{fig2}
\caption{Additional diagrams in coherent state basis for $O(e^2)$ self energy correction corresponding to $T_2$}
\end{figure}
In the coherent state basis, $O(e^2)$ self energy contribution is given by $T^{(1)}+T^{^\prime(1)}$, where $T^{(1)}$ is defined in Eq.~(\ref{def diag1}) and $T^{\prime(1)}$ arises from $O(e^2)$ term in 
\begin{align}\label{coh}
T^\prime(p,p)=&\langle p,s \colon f(p) \vert V_1\vert p,s \colon f(p) \rangle
\end{align}
$\vert p,s \colon f(p) \rangle$ being the coherent state given by Eq.~(\ref{state1}). The contribution of $O(e)$ term in $f(p)$ leads to additional diagrams shown in Fig. 2(a) and Fig. 2(b) and is denoted by $T_{2a}^\prime$+$T_{2b}^\prime$. Here a dotted line represents the soft photon in the coherent state. The diagrams in Fig. 2 correspond to emission (absorption) of soft photon by the incoming(outgoing) fermion, but since the emitted (absorbed) photon is soft, the two particle state containing it is indistinguishable from a single fermion state.
Substituting for $V_1$ from Eq.~(\ref{V_1}) we obtain,
\begin{align}\label{T1}
T^\prime(p,p)=&\frac{e}{(2\pi)^{3/2}}\int{[d\overline p_1][dp_1][dk_1]}\frac{\overline u(\overline p_1,s^\prime _1) \epsilon\llap/^\lambda(k_1)u(p_1,s_1)}{\sqrt{2\overline {p_1^+}}\sqrt{2p_1^+}\sqrt{2k_1^+}}\nonumber\\ &\biggl[\langle p,s \colon f(p) \vert b^\dagger(\overline p_1,s^\prime_1)b(p_1,s_1)a(k_1,\lambda)\vert p,s \colon f(p) \rangle \delta^3(\overline p_1-p_1-k_1)\nonumber\\&+\langle p,s \colon f(p) \vert b^\dagger(\overline p_1,s^\prime _1) b(p_1,s_1)a^\dagger(k_1,\lambda)\vert p,s \colon f(p) \rangle \delta^3(\overline p_1-p_1+k_1)\biggr]
\end{align}
Using coherent state properties Eqs.~(\ref{eigen1}) and (\ref{eigen3}) from Appendix B in Eq.~(\ref{T1}) we obtain 
\begin{align}
T^\prime(p,p)=\frac{e^2}{(2\pi)^{3}}\int\frac{d^2{\bf k}_{1\perp}}{2p^+}\int \frac{dk_1^+}{2k_1^+}\overline u (\overline p,s^\prime) \epsilon\llap/^\lambda(k_1)u(p,s)f(k_1,\lambda:p)
\end{align}
Using Eqs.~(\ref{deltam}), (\ref{uubar}) and calculating the trace,  we obtain
\begin{align}\label{diag2}
{(\delta m^2)}^\prime=\frac{e^2}{(2 \pi)^3} \int{d^2{\bf k}_{1\perp}}\int\frac{dk_1^+}{k_1^+}\frac{(p\cdot \epsilon(k_1))^2 \Theta_\Delta (k_1)}{p\cdot k_1}
\end{align}
where the prime indicates the correction due to additional terms in coherent state basis. \\
The energy denominator in Eq.~(\ref{diag2}) vanishes in the limit $k_1^+ \rightarrow 0,{\bf k}_{1\perp}\rightarrow 0$ thus leading to IR divergences. However, adding  Eqs.~(\ref{diag1}) and (\ref{diag2}), these true IR divergences get cancelled and the $O(e^2)$ electron mass correction is IR divergence free.
\vskip 0.5cm
\section{ELECTRON MASS RENORMALIZATION UPTO $O(e^4)$ IN FOCK BASIS}

We will now calculate $O(e^4)$ electron mass correction in Fock basis. Transition matrix element for $O(e^4)$ correction to self energy is given by
\begin{eqnarray}
T^{(2)}=T_{3}+T_{4}+T_{5}+T_{6}+T_{7}                       
\end{eqnarray}
where
\begin{align}\label{T_3}
T_{3}=&\langle p,s \vert V_1 \frac{1} {p^- - H_0}V_1 \frac{1} {p^- - H_0}V_1 \frac{1} {p^- - H_0}V_1 \vert p,s \rangle \\ \label{T_4}
T_{4}=&\langle p,s \vert V_1\frac{1} {p^- - H_0}V_1\frac{1} {p^- - H_0}V_2\vert p,s \rangle\\ \label{T_5}
T_{5}=&\langle p,s \vert V_1\frac{1} {p^- - H_0}V_2\frac{1} {p^- - H_0}V_1\vert p,s \rangle\\ \label{T_6}
T_{6}=&\langle p,s \vert V_2\frac{1} {p^- - H_0}V_1\frac{1} {p^- - H_0}V_1\vert p,s \rangle\\
\label{T_7} 
T_{7}=&\langle p,s \vert V_2\frac{1} {p^- - H_0}V_2\vert p,s \rangle
\end{align}
These matrix elements correspond to Figs. 3-7 and can be evaluated in the standard manner by inserting appropriate number of complete sets of intermediate states. 
We give the details of the calculation in Appendix C and present here the results.

\begin{figure}[h]
\includegraphics[scale=0.4]{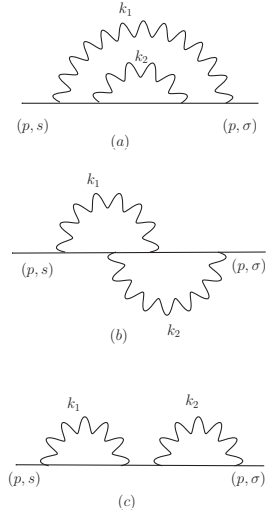}
\label{fig3}
\caption{Diagrams for $O(e^4)$ self energy correction in fock basis corresponding to $T_3$ }
\end{figure}
$T_3$ is given by
\begin{align}
T_{3}\equiv T_3(p,p)=T_{3a}+T_{3b}+T_{3c}
\end{align}
where $T_{3a}$, $T_{3b}$ and $T_{3c}$ correspond to Figs. 3(a)-(c) and are given by Eqs.~(\ref{T_{3a}}), (\ref{T_{3b}}) and (\ref{T_{3c}}) respectively. Using expressions for energy denominator in Eqs.~(\ref{relation1}) and (\ref{relation2}) and using Eq.~(\ref{deltam}) we obtain
\begin{align}
(\delta m^2)_{3a}=&-\frac{e^4}{2(2\pi)^6} \int {{d^2 {\bf k}_{1\perp}}{d^2{\bf k}_{2\perp}}}\int {{{dk_1^+}\over{k_1^+}}{{dk_2^+}\over {k_2^+}}}\nonumber\\ &\frac{Tr[\not\epsilon^{\lambda_1} (k_1)(\not p_1+m)\not\epsilon^{\lambda_2}(k_2)(\not p_2+m) \not\epsilon^{\lambda_2} (k_2)(\not p_1+m) \not\epsilon^{\lambda_1}(k_1)(\not p+m)]}{32(p\cdot k_1)^2[(p\cdot k_1)+(p\cdot k_2)-(k_1\cdot k_2)]}
\end{align}
\begin{align}
(\delta m^2)_{3b}=&-\frac{e^4}{2(2\pi)^6} \int {{d^2 {\bf k}_{1\perp}}{d^2{\bf k}_{2\perp}} }\int {{{dk_1^+}\over{k_1^+}}{{dk_2^+}\over {k_2^+}}}\nonumber\\ &\frac{Tr[\not\epsilon^{\lambda_2} (k_2)(\not p_3+m)\not\epsilon^{\lambda_1}(k_1)(\not p_2+m)\not\epsilon^{\lambda_2}(k_2)(\not p_1+m) \not\epsilon^{\lambda_1}(k_1)(\not p+m)]}{32(p\cdot k_1)(p\cdot k_2)[(p\cdot k_1)+(p\cdot k_2)-(k_1\cdot k_2)]}
\end{align}
\begin{align}
(\delta m^2)_{3c}=&\frac{e^4}{2(2\pi)^6} \int {{d^2 {\bf k}_{1\perp}}{d^2{\bf k}_{2\perp}} }\int {{{dk_1^+}\over{k_1^+}}{{dk_2^+}\over {p^+k_2^+}}}\nonumber\\ &\frac{Tr[\not\epsilon^{\lambda_2} (k_2)(\not p_3+m)\not\epsilon^{\lambda_2}(k_2)(\not p^{\prime}_2+m)\not\epsilon^{\lambda_1}(k_1)(\not p_1+m) \not\epsilon^{\lambda_1}(k_1)(\not p+m)]}{32(p\cdot k_1)(p\cdot k_2)(p^--p^{\prime-}_2)}
\end{align}
where $p_2^\prime =p$ and $p_1$, $p_2$ and $p_3$ have been defined in Eqs.~(\ref{p1}), (\ref{p2}) and (\ref{p3}) respectively.

Note that $(\delta m^2)_{3a}$, $(\delta m^2)_{3b}$ and $(\delta m^2)_{3c}$ can have IR divergences when 
\begin{description}
\item[I]   $p\cdot k_1 \rightarrow 0$ i.e $k_1^+\rightarrow 0, {\bf k}_{1\perp}\rightarrow 0$, but $p\cdot k_2 \neq 0$.	\item[II]  $p\cdot k_2\rightarrow 0$ i.e $k_2^+\rightarrow 0, {\bf k}_{2\perp}\rightarrow 0$, but $p\cdot k_1 \neq 0$.
\item[III] $p\cdot k_1 \rightarrow 0$ and $p\cdot k_2 \rightarrow 0$ i.e. $k_1^+\rightarrow 0, {\bf k}_{1\perp}\rightarrow 0$, $k_2^+\rightarrow 0, {\bf k}_{2\perp}\rightarrow 0$.
\end{description}
Now we will consider the contribution of Figs. 3(a)-(c) in each of these limits.\\
Case I : In the  limit  $k_1^+\rightarrow 0, {\bf k}_{1\perp}\rightarrow 0$ (but $ p\cdot k_2 \not\rightarrow 0$), the contribution to $T_3$ from diagrams in Figs. 3(a) and 3(b) is given by
\begin{align}\label{3aI}
&[(\delta m^2)_{3a}+(\delta m^2)_{3b}]^{I}\nonumber\\
&=-{e^4\over{(2\pi)^6}}\int{{d^2{\bf k}_{1\perp}}{d^2{\bf k}_{2\perp}}}\int{{{dk_1^+}\over{k_1^+}}{{dk_2^+}\over {k_2^+}}}\frac{[2(p\cdot\epsilon(k_1))^2(p\cdot \epsilon(k_2))^2-(p\cdot k_2)(p\cdot \epsilon(k_1))^2]}{4(p\cdot k_1)^2(p\cdot k_2)}
\end{align}
and contribution from Fig. 3(c) is given by
\begin{align}\label{3cI}
[(\delta m^2)_{3c}]^{I}=&{e^4\over{(2 \pi)^6}} \int {{d^2{\bf k}_{1\perp}}{d^2{\bf k}_{2\perp}} }\int {{{dk_1^+}\over{k_1^+}}{{dk_2^+}\over {k_2^+}}}\frac{2(p\cdot \epsilon(k_1))^2(p\cdot \epsilon(k_2))^2-(p\cdot k_2)(p\cdot \epsilon(k_1))^2}{8p^+}\nonumber\\&\biggl[\frac{p^+}{(p\cdot k_1)^2(p\cdot k_2)}+\frac{p_3^+}{(p\cdot k_1)(p\cdot k_2)^2}\biggr]
\end{align}
Here we have used Heitler method \cite{HEIT54} illustrated in Appendix C to deal with the vanishing denominator $(p^--p_2^{\prime-})$. \\
Case II :  In the  limit  $k_2^+\rightarrow 0, {\bf k}_{2\perp}\rightarrow 0$ (but $ p\cdot k_1 \not\rightarrow 0$), the contribution to $T_3$ from diagrams in Figs. 3(a) and 3(b) is given by
\begin{align}\label{3aII}
&[(\delta m^2)_{3a}+(\delta m^2)_{3b}]^{II}\nonumber\\&=-{e^4\over{(2\pi)^6}}\int{{d^2{\bf k}_{1\perp}}{d^2{\bf k}_{2\perp}}}\int{{{dk_1^+}\over{k_1^+}}{{dk_2^+}\over{k_2^+}}}\frac{[2(p\cdot \epsilon(k_1))^2(p\cdot \epsilon(k_2))^2-(p\cdot k_1)(p\cdot\epsilon(k_2))^2]}{4(p\cdot k_1)(p\cdot k_2)^2}
\end{align}
and contribution from Fig. 3(c) is given by
\begin{align}\label{3cII}
[(\delta m^2)_{3c}]^{II}=&{e^4\over{(2\pi)^6}}\int{{d^2{\bf k}_{1\perp}}{d^2{\bf k}_{2\perp}}}\int{{{dk_1^+}\over {k_1^+}}{{dk_2^+}\over{k_2^+}}}\frac{2(p\cdot \epsilon(k_1))^2(p\cdot \epsilon(k_2))^2-(p\cdot k_1)(p\cdot \epsilon(k_2))^2}{8p^+}\nonumber\\&\biggl[\frac{p_1^+}{(p\cdot k_1)^2(p\cdot k_2)}+\frac{p^+}{(p\cdot k_1)(p\cdot k_2)^2}\biggr]
\end{align}
Case III : In the  limit  $k_1^+\rightarrow 0, {\bf k}_{1\perp}\rightarrow 0, k_2^+\rightarrow 0, {\bf k}_{2\perp}\rightarrow 0$, the contribution to $T_3$ from diagrams in Figs. 3(a) and 3(b) is given by
\begin{align}\label{3aIII}
[(\delta m^2)_{3a}+(\delta m^2)_{3b}]^{III}=-{e^4\over{(2 \pi)^6}}\int{{d^2{\bf k}_{1\perp}}{d^2{\bf k}_{2\perp}}}\int {{{dk_1^+}\over{k_1^+}}{{dk_2^+}\over{k_2^+}}}\frac{(p\cdot\epsilon(k_1))^2(p\cdot\epsilon(k_2))^2}{2(p\cdot k_1)^2 (p\cdot k_2)}
\end{align}
and contribution from Fig. 3(c) is given by
\begin{align}\label{3cIII}
[(\delta m^2)_{3c}]^{III}=&{e^4\over{(2 \pi)^6}}\int {{d^2{\bf k}_{1\perp}}{d^2{\bf k}_{2\perp}}}\int{{{dk_1^+} \over{k_1^+}}{{dk_2^+}\over{k_2^+}}}\biggl[\frac{(p\cdot \epsilon(k_1))^2(p\cdot \epsilon(k_2))^2}{4(p\cdot k_1)^2(p\cdot k_2)}\nonumber\\&+\frac{(p\cdot \epsilon(k_1))^2(p\cdot \epsilon(k_2))^2}{4(p\cdot k_1)(p\cdot k_2)^2}\biggr]
\end{align}

\begin{figure}[h]
\includegraphics[scale=0.4]{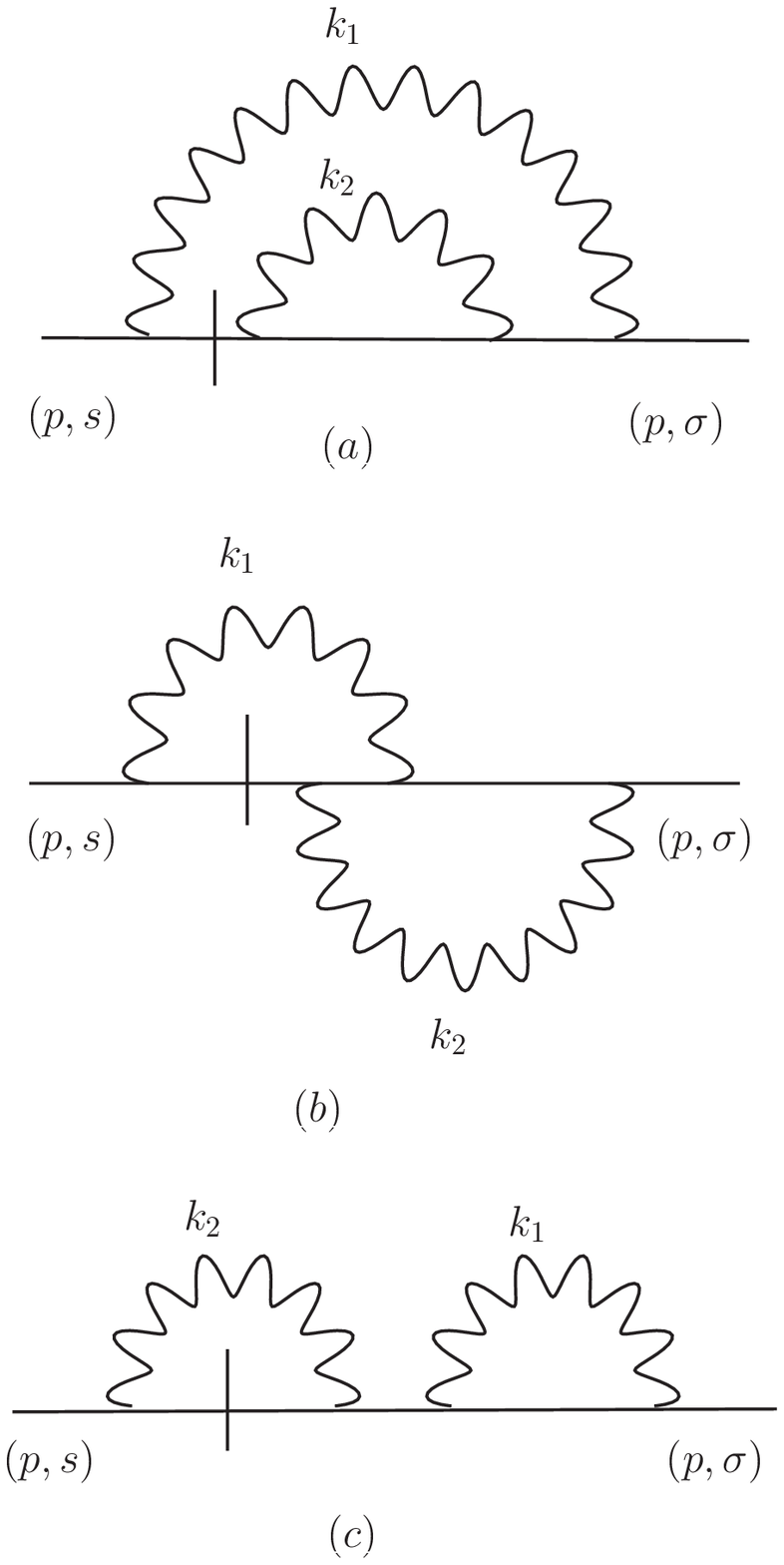}
\label{fig4}
\caption{Diagrams for $O(e^4)$ self energy correction in fock basis corresponding to $T_4$ }
\end{figure}
\begin{figure}
\includegraphics[scale=0.4]{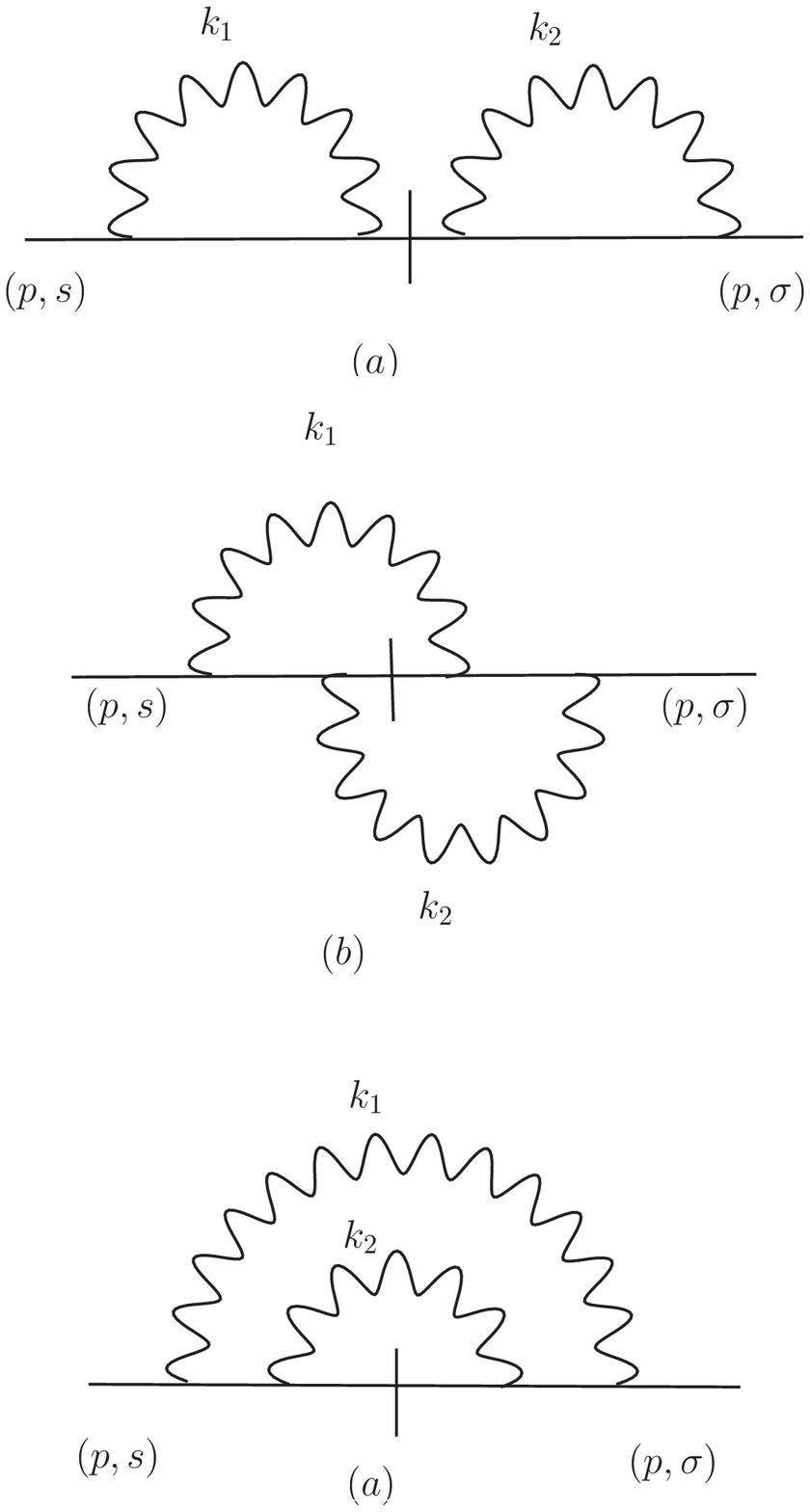}
\label{fig5}
\caption{Diagrams for $O(e^4)$ self energy correction in fock basis corresponding to $T_5$ }
\end{figure}

\begin{figure}
\includegraphics[scale=0.4]{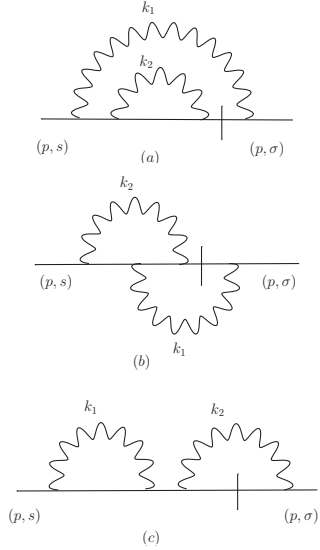}
\label{fig6}
\caption{Diagrams for $O(e^4)$ self energy correction in fock basis corresponding to $T_6$ }
\end{figure}


The contributions to $T_4$, $T_5$ and $T_6$ come from diagrams in Figs. 4, 5 and 6 respectively.
One can do similar calculation for the three cases by carefully taking the appropriate limits of corresponding expressions. Below we give contributions from these diagrams in each of the three limits. 

Case I : In the limit $k_1^+\rightarrow 0, {\bf k}_{1\perp}\rightarrow 0$ (but $ p\cdot k_2 \not\rightarrow 0$), the contributions of diagrams in Figs. 4(a), 5(a), 5(b) and 6(a) are given by
\begin{align}\label{4aI}
[(\delta m^2)_{4a}]^{I}=[(\delta m^2)_{6a}]^{I}=&-{e^4\over{(2\pi)^6}}\int{{d^2{\bf k}_{1\perp}}{d^2{\bf k}_{2\perp}}}\int{{{dk_1^+}\over {k_1^+}}{{dk_2^+}\over{k_2^+p^+}}}\nonumber\\ &\frac{[2p^+(p\cdot \epsilon(k_1))(p\cdot \epsilon(k_2))(\epsilon(k_1) \cdot\epsilon(k_2))+k_2^+(p\cdot\epsilon(k_1))^2]}{8(p\cdot k_1)(p\cdot k_2)}
\end{align}
\begin{align}\label{5aI}
[(\delta m^2)_{5a}]^{I}=&-{e^4\over{(2\pi)^6}}\int{{d^2{\bf k}_{1\perp}}{d^2{\bf k}_{2\perp}}}\int{{{dk_1^+}\over {k_1^+}}{{dk_2^+}\over{k_2^+p^+}}}\nonumber\\ &\frac{[2p^+(p\cdot \epsilon(k_1))(p\cdot \epsilon(k_2)) (\epsilon(k_1)\cdot\epsilon(k_2))+k_2^+(p\cdot\epsilon(k_1))^2]}{8(p\cdot k_1)(p\cdot k_2)}
\end{align}
\begin{align}\label{5bI}
[(\delta m^2)_{5b}]^{I}=&-{e^4\over{(2\pi)^6}}\int{{d^2{\bf k}_{1\perp}}{d^2{\bf k}_{2\perp}}}\int{{{dk_1^+}\over {k_1^+}}{{dk_2^+}\over{k_2^+p_3^+}}}\nonumber\\ &\frac{[2p_3^+(p\cdot\epsilon(k_1))(p\cdot\epsilon(k_2))(\epsilon(k_1)\cdot \epsilon(k_2))-k_2^+(p\cdot \epsilon(k_1))^2]}{8(p\cdot k_1)(p\cdot k_2)}
\end{align}
Diagrams in Figs. 4(c), 5(c) and 6(c) have IR divergences only in limit I and lead to 
\begin{align}\label{4cI} 
[(\delta m^2)_{4c}+(\delta m^2)_{6c}]^{I}=-{e^4\over{(2\pi)^6}}\int{{d^2{\bf k}_{1\perp}}{d^2{\bf k}_{2\perp}}}\int {{{dk_1^+}\over{k_1^+}}{{dk_2^+}\over{k_2^+}}}\frac{{(p\cdot\epsilon(k_1))}^2}{{2(p\cdot k_1)}^2}
\end{align}
\begin{align}\label{5cI}
[(\delta m^2)_{5c}]^{I}={e^4\over{(2\pi)^6}}\int{{d^2{\bf k}_{1\perp}}{d^2{\bf k}_{2\perp}}}\int{{{dk_1^+}\over {k_1^+}}{{dk_2^+}\over{k_2^+}}}\frac{{(p\cdot\epsilon(k_1))}^2}{{2(p\cdot k_1)}^2}
\end{align}
where again we have used Heitler method for evaluating $T_{4c}+T_{6c}$. Figs. 4(b) and 6(b) do not have IR divergences in this limit.\\
Case II : In the limit $k_2^+\rightarrow 0, {\bf k}_{2\perp}\rightarrow 0$ (but $ p\cdot k_1 \not\rightarrow 0$), the contributions of diagrams in Figs. 4(b), 5(a), 5(b) and 6(b) are given by following expressions
\begin{align}\label{4bII}
[(\delta m^2)_{4b}]^{II}=[(\delta m^2)_{6b}]^{II}=&-{e^4\over{(2\pi)^6}}\int{{d^2{\bf k}_{1\perp}}{d^2{\bf k}_{2\perp}}}\int{{{dk_1^+}\over {k_1^+}}{{dk_2^+}\over{k_2^+p_1^+}}}\nonumber\\ &\frac{[2p_1^+(p\cdot \epsilon(k_1))(p\cdot \epsilon(k_2))(\epsilon(k_1)\cdot \epsilon(k_2))-k_1^+(p\cdot\epsilon(k_2))^2]}{8(p\cdot k_1)(p\cdot k_2)}
\end{align}
\begin{align}\label{5aII}
[(\delta m^2)_{5a}]^{II}=&-{e^4\over{(2\pi)^6}}\int{{d^2{\bf k}_{1\perp}}{d^2{\bf k}_{2\perp}}}\int{{{dk_1^+}\over {k_1^+}}{{dk_2^+}\over{k_2^+p_1^+}}}\nonumber\\ &\frac{[2p^+(p\cdot\epsilon(k_1))(p\cdot \epsilon(k_2)) (\epsilon(k_1)\cdot\epsilon(k_2))+k_1^+(p\cdot\epsilon(k_2))^2]}{8(p\cdot k_1)(p\cdot k_2)}
\end{align}
\begin{align}\label{5bII}
[(\delta m^2)_{5b}]^{II}=&-{e^4\over{(2\pi)^6}}\int{{d^2{\bf k}_{1\perp}}{d^2{\bf k}_{2\perp}}}\int{{{dk_1^+}\over {k_1^+}}{{dk_2^+}\over{k_2^+p_1^+}}}\nonumber\\ &\frac{[2p_1^+(p\cdot\epsilon(k_1))(p\cdot\epsilon(k_2)) (\epsilon(k_1)\cdot\epsilon(k_2))-k_1^+(p\cdot\epsilon(k_2))^2]}{8(p\cdot k_1)(p\cdot k_2)}
\end{align}
Figs. 4(a) and 6(a) do not have IR divergences in this limit. \\
Case III : In the  limit  $k_1^+\rightarrow 0, {\bf k}_{1\perp}\rightarrow 0, k_2^+\rightarrow 0, {\bf k}_{2\perp}\rightarrow 0$, the sum of contributions corresponding to Figs. 4-6(a) and (b) is given by
\begin{align}\label{4aIII}
[{(\delta m^2)}_{4a}+{(\delta m^2)}_{4b}+{(\delta m^2)}_{5a}+{(\delta m^2)}_{5b}+{(\delta m^2)}_{6a}+{(\delta m^2)}_{6b}]^{III}\nonumber\\
=-{e^4\over{(2\pi)^6}}\int {{d^2{\bf k}_{1\perp}}{d^2{\bf k}_{2\perp}}}\int{{{dk_1^+}\over{k_1^+}}{{dk_2^+}\over {k_2^+}}}\frac{[(p\cdot\epsilon(k_1))(p\cdot\epsilon(k_2))(\epsilon(k_1)\cdot\epsilon(k_2))]}{(p\cdot k_1)(p\cdot k_2)}
\end{align}
where we have used Eqs.~(\ref{4a3}), (\ref{5a3}), (\ref{5b3}) and (\ref{6a3}). 
\begin{figure}[h]
\includegraphics[scale=0.4]{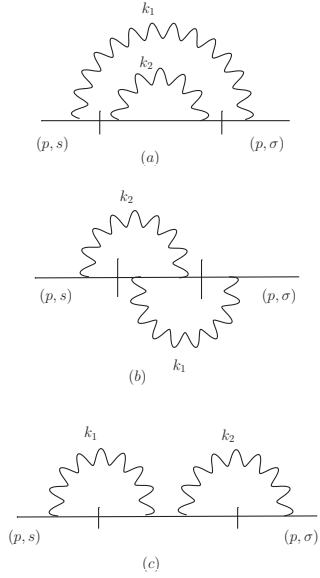}
\label{fig7}
\caption{Diagrams for $O(e^4)$ self energy correction in fock basis corresponding to $T_7$ }
\end{figure}
The last term $T_{7}$ in Eq.~(\ref{T_7}) is IR convergent as the 4-point energy denominator involved here is non-zero and hence $T_{7}$ is not needed for our discussion. One can notice that Eqs.~(\ref{3aI})-(\ref{4aIII}) have true IR divergences since the denominator vanishes as $k^+_1\rightarrow 0$, ${\bf k}_{1\perp}\rightarrow 0$ and/or $k^+_2\rightarrow 0$, ${\bf k}_{2\perp}\rightarrow 0$.
\vskip 0.25cm 
\section{ELECTRON MASS RENORMALIZATION IN COHERENT STATE BASIS UPTO $O(e^4)$}
In this section, we will use coherent state basis to calculate the $O(e^4)$ electron mass correction. We will show that the IR divergences in additional diagrams appearing due to the use of coherent state basis exactly cancel the IR divergences arising due to vanishing energy denominators calculated in Section V. In coherent state basis, $O(e^4)$  correction to self energy is given by
\begin{align}
T^{(2)}+T_8^\prime+T_9^\prime+T_{10}^\prime+T_{11}^\prime\; \nonumber
\end{align}
where $T_8^\prime$ is $O(e^4)$ term in $\langle p,s \colon f(p) \vert V_1 \frac{1} {p^- - H_0}V_1\frac{1} {p^- - H_0}V_1\vert p,s \colon f(p) \rangle$ represented by Fig. 8, $T_9^\prime$ is $O(e^4)$ term in $\langle p,s \colon f(p) \vert V_1 \frac{1} {p^- - H_0}V_1\vert p,s \colon f(p) \rangle$ represented by Fig. 9, $T_{10}^\prime$ is $O(e^4)$ term in $\langle p,s\colon f(p)\vert V_1\frac{1} {p^- - H_0}V_2\vert p,s \colon f(p)\rangle+\langle p,s\colon f(p)\vert V_2 \frac{1} {p^- - H_0}V_1\vert p,s \colon f(p)\rangle$ represented by Fig. 10 and $T_{11}^\prime$ is $O(e^4)$ term in $\langle p,s\colon f(p)\vert V_2 \vert p,s \colon f(p)\rangle$ represented by Fig. 11. We present the details of calculation in Appendix D and give below  only the result for $(\delta m^2)^\prime$. 

\begin{figure}[h]
\includegraphics[scale=0.4]{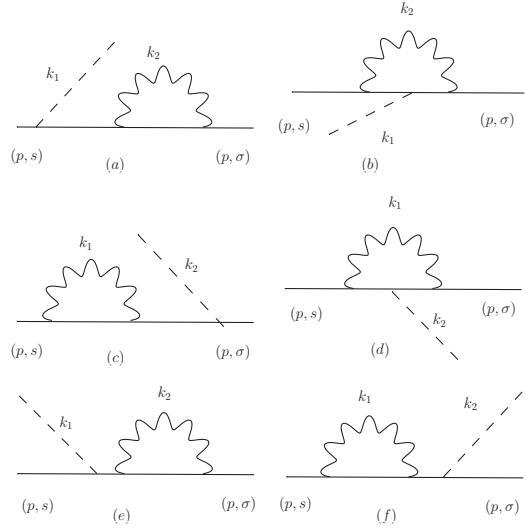}
\label{fig8}
\caption{Additional diagrams in coherent state basis for $O(e^4)$ self energy correction corresponding to $T_8$ }
\end{figure}
The contribution corresponding to Fig. 8 is given by
\begin{equation}
{(\delta m^2)}_{8}^\prime={(\delta m^2)}_{8a}^\prime+{(\delta m^2)}_{8b}^\prime+{(\delta m^2)}_{8c}^\prime+{(\delta m^2)}_{8d}^\prime+{(\delta m^2)}_{8e}^\prime+{(\delta m^2)}_{8f}^\prime
\end{equation}
where $(\delta m^2)_{8a}^\prime$ - $(\delta m^2)_{8f}^\prime$ have been evaluated in Appendix D. 

Figs. 8a and 8b contribute 
\begin{align}\label{coh8aI}
{(\delta m^2)}_{8a}^\prime=&\frac{e^4}{(2\pi)^6}\int{{d^2{\bf k}_{1\perp}}{d^2{\bf k}_{2\perp}}}\int{{{dk_1^+} \over{k_1^+}}{{dk_2^+}\over{k_2^+}}}\nonumber\\ &\frac{[2(p\cdot\epsilon(k_1))^2(p\cdot\epsilon(k_2))^2-(p\cdot k_2)(p\cdot \epsilon(k_1))^2]}{4(p\cdot k_1)^2[(p\cdot k_1)+(p\cdot k_2)-(k_1\cdot k_2)]}\Theta_{\Delta}(k_1)
\end{align}
\begin{align}\label{coh8bI}
{(\delta m^2)}_{8b}^\prime=&{e^4\over{(2 \pi)^6}}\int{{d^2{\bf k}_{1\perp}}{d^2{\bf k}_{2\perp}}}\int{{{dk_1^+} \over{k_1^+}}{{dk_2^+}\over{k_2^+}}}\nonumber\\ &\frac{[2(p\cdot\epsilon(k_1))^2(p\cdot\epsilon(k_2))^2]}{4(p\cdot k_1)(p\cdot k_2)[(p\cdot k_1)+(p\cdot k_2)-(k_1\cdot k_2)]}\Theta_{\Delta}(k_1)
\end{align}
One can notice that for $(\delta m^2)_{8a}^\prime$ and $(\delta m^2)_{8b}^\prime$ we need not discuss limit II as $p\cdot k_1$ is always small. Adding Eqs.~(\ref{coh8aI}) and (\ref{coh8bI}) and taking the limit I, we obtain
\begin{align}\label{coh8a}
[(\delta m^2)_{8a}^\prime+(\delta m^2)_{8b}^\prime]^I=&{e^4\over{(2 \pi)^6}}\int{{d^2{\bf k}_{1\perp}}{d^2{\bf k}
_{2\perp}}}\int{{{dk_1^+}\over{k_1^+}}{{dk_2^+}\over{k_2^+}}}\nonumber\\ &\biggl[\frac{(p\cdot\epsilon(k_1))^2(p\cdot\epsilon(k_2))^2}{2(p\cdot k_1)^2(p\cdot k_2)}-\frac{(p\cdot \epsilon(k_1))^2}{4(p\cdot k_1)^2}\biggr]\Theta_{\Delta}(k_1)
\end{align}
\begin{figure}[h]
\includegraphics[scale=0.4]{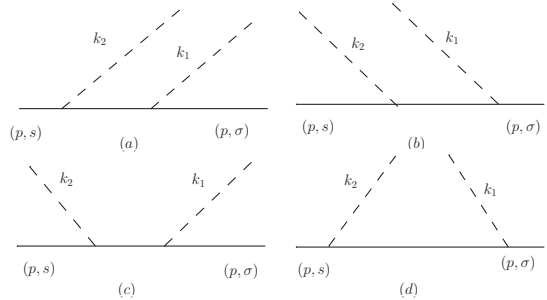}
\label{fig9}
\caption{Additional diagrams in coherent state basis for $O(e^4)$ self energy correction corresponding to $T_9$ }
\end{figure}

Adding  Eqs.~(\ref{3aI}) and (\ref{coh8a}) we find that $[(\delta m^2)_{3a}+{(\delta m^2)}_{3b}]^{I}$+$[(\delta m^2)_{8a}^\prime+{(\delta m^2)}_{8b}^\prime]^{I}$ is IR finite. \\
Adding all the contributions coming from Figs. 3, 8 and 9, we find that the IR divergences completely cancel. Below we summarize this result for the reader's convenience:
\begin{enumerate}
	\item $[(\delta m^2)_{3a}+{(\delta m^2)}_{3b}]^{I}$+$[(\delta m^2)_{8a}^\prime+{(\delta m^2)}_{8b}^\prime]^{I}$ is IR finite.
	\item $[(\delta m^2)_{3a}+{(\delta m^2)}_{3b}]^{II}$+$[(\delta m^2)_{8c}^\prime+{(\delta m^2)}_{8d}^\prime]^{II}$ is IR finite.
	\item $[(\delta m^2)_{3c}]^{I}$+$[(\delta m^2)_{8e}^\prime]^{I}$ is IR finite.
	\item $[(\delta m^2)_{3c}]^{II}$+$[(\delta m^2)_{8f}^\prime]^{II}$ is IR finite.
	\item $[(\delta m^2)_{3}]^{III}+[(\delta m^2)_{8a}^\prime+(\delta m^2)_{8b}^\prime]^{III}+[(\delta m^2)_{8e}^\prime]^{III}$ is IR finite.
	\item $[(\delta m^2)_{8c}^\prime+(\delta m^2)_{8d}^\prime]^{III}+[(\delta m^2)_{8f}^\prime]^{III}+[(\delta m^2)_{9}^\prime]^{III}$ is IR finite.	
\end{enumerate}
Thus, we can see that the self energy correction corresponding to 3-point QED vertices upto $O(e^4)$ is IR finite. In the same manner we can show the cancellation of IR divergences for diagrams containing 4-point instantaneous fermion exchange vertex in all the three limits.
\begin{figure}[h]
\includegraphics[scale=0.6]{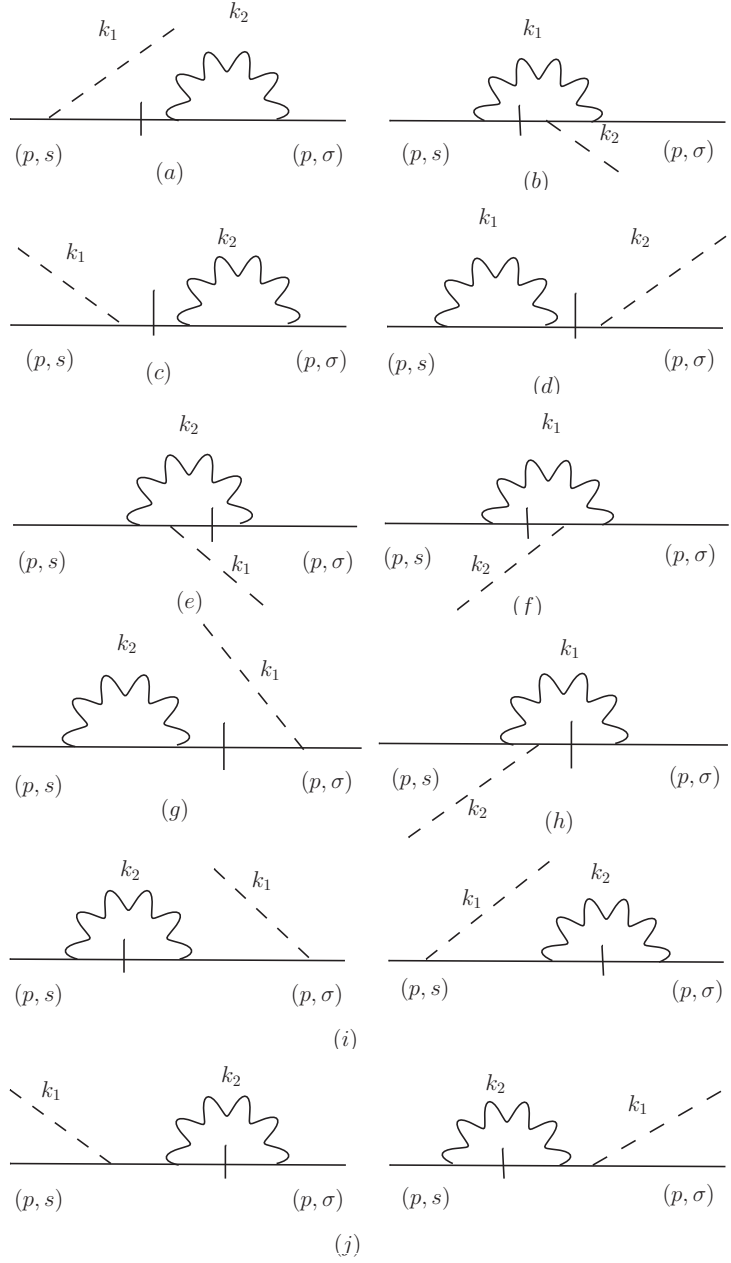}
\label{fig10}
\caption{Additional diagrams in coherent state basis for $O(e^4)$ self energy correction corresponding to $T_{10}$ }
\end{figure}
We give here one calculation for illustration.
The contributions corresponding to Fig. 10 is given by
\begin{align}
(\delta m^2)_{10}^\prime=(\delta m^2)_{10a}^\prime+(\delta m^2)_{10b}^\prime+(\delta m^2)_{10c}^\prime+(\delta m^2)_{10d}^\prime+(\delta m^2)_{10e}^\prime\nonumber\\+(\delta m^2)_{10f}^\prime+(\delta m^2)_{10g}^\prime+(\delta m^2)_{10h}^ \prime+(\delta m^2)_{10i}^\prime+(\delta m^2)_{10j}^\prime
\end{align}
As shown in Appendix D, the IR divergent contributions in Fig. 10(a), 10(c), 10(e) and 10(g) in the limit I are given by the following expressions:
\begin{align}\label{coh10a}
[(\delta m^2)_{10a}^\prime]^{I}=[(\delta m^2)_{10g}^\prime]^{I}=&\frac{e^4}{(2\pi)^6}\int{{d^2{\bf k}_{1\perp}}{d^2{\bf k}_{2\perp}}}\int{{{dk_1^+} \over{k_1^+}}{{dk_2^+}\over{k_2^+}}}\nonumber\\ &\frac{[2p^+(p\cdot\epsilon(k_1))(p\cdot \epsilon(k_2)) (\epsilon(k_1)\cdot\epsilon(k_2))+k_2^+(p\cdot \epsilon(k_1))^2]}{8p^+(p\cdot k_1)(p\cdot k_2)}\Theta_{\Delta}(k_1)
\end{align} 
\begin{align}\label{coh10c}
[(\delta m^2)_{10c}^\prime]^{I}=&{e^4\over{(2\pi)^6}}\int{{d^2{\bf k}_{1\perp}}{d^2{\bf k}_{2\perp}}}\int{{{dk_1^+} \over{k_1^+}}{{dk_2^+}\over{k_2^+}}}\nonumber\\&\frac{[2p^+(p\cdot \epsilon(k_1))(p\cdot \epsilon(k_2)) (\epsilon(k_1)\cdot\epsilon(k_2))+k_2^+(p\cdot \epsilon(k_1))^2]}{8p^+(p\cdot k_1)(p\cdot k_2)}\Theta_{\Delta}(k_1)
\end{align} 
\begin{align}\label{coh10e}
[(\delta m^2)_{10e}^\prime]^{I}=&{e^4\over{(2\pi)^6}}\int{{d^2{\bf k}_{1\perp}}{d^2{\bf k}_{2\perp}}}\int{{{dk_1^+} \over{k_1^+}}{{dk_2^+}\over{k_2^+}}}\nonumber\\&\frac{[2p_3^+(p\cdot \epsilon(k_1))(p\cdot \epsilon(k_2)) (\epsilon(k_1)\cdot\epsilon(k_2))-k_2^+(p\cdot \epsilon(k_1))^2]}{8p_3^+(p\cdot k_1)(p\cdot k_2)}\Theta_{\Delta}(k_1)
\end{align} 
\begin{figure}[h]
\includegraphics[scale=0.6]{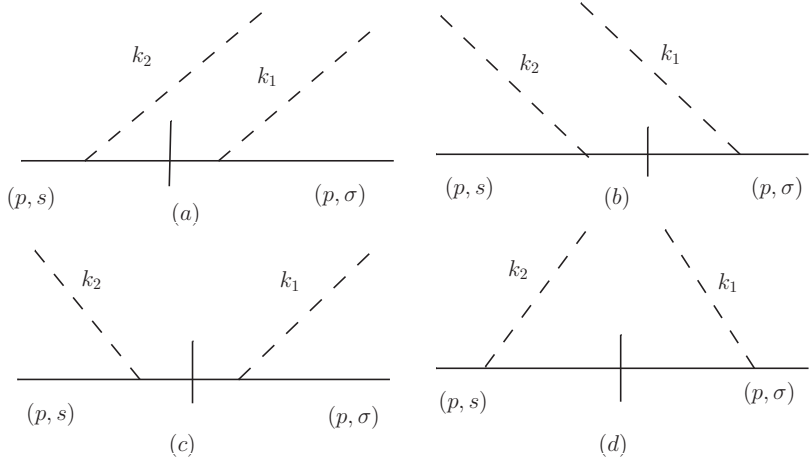}
\label{fig11}
\caption{Additional diagrams in coherent state basis for $O(e^4)$ self energy correction corresponding to $T_{11}$}
\end{figure}\\
Adding Eqs.~(\ref{4aI}-\ref{5bI}) and Eqs.~(\ref{coh10a} -\ref{coh10e}),  we find that $[(\delta m^2)_{4a}]^{I}$+$[(\delta m^2)_{5a}]^{I}$+$[(\delta m^2)_{5b}]^{I}$+$[(\delta^2)_{6a}]^{I}$+$[(\delta m^2)_{10a}^\prime]^{I}$+$[(\delta m^2)_{10c}^\prime]^{I}$+$[(\delta m^2)_{10e}^\prime]^{I}$ \\
+$[(\delta m^2)_{10g} ^\prime]^{I}$ is IR finite. 
Adding all the contributions from Figs. 4, 5, 6, 10 and 11, we find that the IR divergences exactly cancel. Below we summarize our results:
\begin{enumerate}
	\item$[(\delta m^2)_{4a}]^{I}$+$[(\delta m^2)_{5a}]^{I}$+$[(\delta m^2)_{5b}]^{I}$+$[(\delta^2)_{6a}]^{I}$+$[(\delta m^2)_{10a}^\prime]^{I}$+$[(\delta m^2)_{10c}^\prime]^{I}$+$[(\delta m^2)_{10e}^\prime]^{I}$ \\
+$[(\delta m^2)_{10g} ^\prime]^{I}$ is IR finite. 
	\item $[(\delta m^2)_{4b}]^{II}$+$[(\delta m^2)_{5a}]^{II}$+$[(\delta m^2)_{5b}]^{II}$+$[(\delta m^2)_{6b}]^{II}$+$[(\delta m^2)_{10b}^\prime]^{II}$\nonumber\\ +$[(\delta m^2)_{10d}^\prime]^{II}$+$[(\delta m^2)_{10f}^\prime]^{II}$+$[(\delta m^2)_{10h}^\prime]^{II}$ is IR finite. 
		\item $[(\delta m^2)_{4c}]^{I}+[(\delta m^2)_{5c}^\prime]^{I}+[(\delta m^2)_{6c}^\prime]^{I}+[(\delta m^2)_{10i}^\prime]^{I}+[(\delta m^2)_{10j}^\prime]^{I}$ is IR finite.
	\item $[(\delta m^2)_{4}+(\delta m^2)_{5}+(\delta m^2)_{6}+(\delta m^2)_{10}^\prime+(\delta m^2)_{11}^\prime]^{III}$ is IR finite.
\end{enumerate}
Thus, finally we obtain
\[{(\delta m^2)}^{(2)}+ \sum_{i=8}^{11}{(\delta m^2)}_{i}^{\prime} \]is IR finite, 
where ${(\delta m^2)}^{(2)}$ is $O(e^4)$ electron mass correction  in Fock basis.
This completes the proof of cancellation of true IR divergences upto $O(e^4)$ for fermion self energy correction in coherent state basis.
\vskip 0.5cm
\section{CONCLUSION}
\vskip 0.5cm
We have calculated  electron self energy correction in light-front QED up to $O(e^4)$ and have shown that the {\it true}  IR divergences get cancelled when coherent state basis is used to calculate the matrix elements.  The cancellation of IR divergences between real and virtual processes is known to hold  in equal-time QED to all orders.  This cancellation was also shown by Kulish and Faddeev \cite{KUL70} using the coherent state formalism. It would be interesting to verify this  all order cancellation in  LFQED. The present work is a first step towards this aim. The true IR divergences in QCD do not cancel in higher orders. This fact can possibly be used to obtain a form of the artificial potential required for the bound-state calculation. The connection between the asymptotic dynamics and cancellation/non-cancellation of IR divergences can also be exploited to explore the possibility of constructing an artificial potential which is used in the bound state calculations in LFQCD \cite{WILSON94}. 
\vskip 0.25cm
{\bf ACKNOWLEGEMENTS}
\vskip 0.25cm
We wish to acknowledge the financial support from Department of Science and Technology, India under the Grant No. SR/S2/HEP-17/2006.
\vskip 0.25cm
\appendix 
\vskip .25cm
\section{\it Notation and Useful Relations}
\vskip 0.25cm
We define four-vector $x^{\mu}$ by
\begin{center}
{ $ x^{\mu}=(x^0,x^3,x^1,x^2)=(x^0,x^3,{\bf x}^{\perp}) $}
\end{center}
The  light-front variables are defined by
\begin{equation}
x^+=\frac{(x^0 +x^3)}{\sqrt{2}},\qquad x^-=\frac{(x^0 -x^3)}{\sqrt{2}},\qquad {\bf x}_{\perp}=(x^1,x^2)
\end{equation}
Thus, in light-front variables
\begin{center}
{ $ x^{\mu}=(x^+,x^-,{\bf x}^{\perp}) $}
\end{center}
The metric tensor is
\[
g^{\mu\nu} =
\left[{\begin{array}{cccc}
0&1&0&0\\
1&0&0&0\\
0&0&-1&0\\
0&0&0&-1\\
\end{array}}\right]
\]
\subsection{\it Dirac spinors:}
$u(p,s)$ and $\overline u(p,s)$ satisfy the usual properties
\begin{eqnarray}\label{dirac}
(\not p -m)u(p,s)=0,\quad (\not p +m)v(p,s)=0
\end{eqnarray}
\begin{eqnarray}\label{uu}
\overline u(p,s)u(p,s^\prime)=-\overline v(p,s)v(p,s^\prime)=2m \delta_{ss^\prime}\;,\overline u(p,s)\gamma^{\mu} u(p,s^\prime)=\overline v(p,s)\gamma^{\mu}v(p,s^\prime)=2 p^{\mu}\delta_{ss^\prime}\;,
\end{eqnarray}
\begin{eqnarray}\label{uubar}
\sum_{s=\pm 1/2} u(p,s)\overline{u}(p,s)=\not p+ m \;,\quad \sum_{s=\pm 1/2} v(p,s)\overline{v}(p,s)=\not p- m \;.
\end{eqnarray}
\subsection{\it Photon polarizations:}
Photon polarization tensor $\epsilon_\mu^\lambda$ satisfies
\begin{eqnarray}\label{pol1}
d_{\mu\nu}(p) = \sum_{\lambda=1,2} \epsilon^\lambda_\mu(p)\epsilon^\lambda_\nu(p)
=-g_{\mu\nu} + {{\delta_{\mu+}p_\nu+\delta_{\nu+}p_\mu} \over p^+}\; ,
\end{eqnarray}
Some useful properties satisfied by $d_{\alpha\beta}(p)$ are
\begin{align}
\gamma^{\alpha}\gamma^{\beta}d_{\alpha\beta}(p)=&-2,\\
\gamma^{\alpha}\gamma^{\nu}\gamma^{\beta}d_{\alpha\beta}(p)=&\frac{2}{p^+}(\gamma^+p^{\nu}+g^{+\nu} \not p),\\
\gamma^{\alpha}\gamma^{\mu}\gamma^{\nu}\gamma^{\beta}d_{\alpha\beta}(p)=&-4g^{\mu \nu}+ \frac{2p_\alpha}{p^+}(g^{\mu\alpha}\gamma^\nu \gamma^+-g^{\alpha\nu}\gamma^\mu \gamma^++g^{\alpha+} \gamma^\mu \gamma^\nu-g^{+\nu} \gamma^\mu \gamma^\alpha+g^{+\mu} \gamma^\nu \gamma^\alpha).
\end{align}
\vskip 0.25cm
\subsection{\it Energy denominators:}
We will need the following expressions for energy denominators:
\begin{align}\label{relation1}
p^--k_1^--(p-k_1)^-=&-\frac{(p\cdot k_1)}{p^+-k_1^+}\\
\label{relation2}
p^- - k_1^-- k_2^- -(p-k_1-k_2)^- =&-\frac {p \cdot k_1+p \cdot k_2-k_1 \cdot k_2}{p^+-k_1^+-k_2^+}
\nonumber\\
p^- + k_1^-- k_2^- - (p+k_1-k_2)^- =&\frac {p \cdot k_1-p \cdot k_2-k_1 \cdot k_2}{p^++k_1^+-k_2^+}
\nonumber\\
p^- - k_1^-+ k_2^- - (p-k_1+k_2)^- =&-\frac {p \cdot k_1-p \cdot k_2+k_1 \cdot k_2}{p^+-k_1^-+k_2^+}
\nonumber\\
p^- + k_1^-+ k_2^- - (p+k_1+k_2)^- =&\frac {p \cdot k_1+p \cdot k_2+k_1 \cdot k_2}{p^++k_1^-+k_2^+}
\end{align}
\section{Properties of Coherent States}
The coherent state containing a fermion and superposition of infinite number of soft photons is denoted by $\vert 1 \colon p_i \rangle$ and is defined by Eq.~(\ref{state1}). Similarly, the coherent state containing a fermion and a hard photon is denoted by $\vert 2 \colon p_{i}, k_{i} \rangle$.  
The coherent states $\vert 1\colon p_i\rangle$ are the eigenstates of $a(k,\lambda)$ \cite{ANU94}
\begin{equation}\label{eigen1}
a(k,\rho)\vert 1 \colon p_i \rangle =
-{e \over {(2\pi)^{3/2}}}\frac{f(k,\rho \colon p_i) }{\sqrt{2k^+}}
\vert 1 \colon p_i\rangle \;.
\end{equation}
Also,
\begin{equation}\label{eigen2}
a(k,\rho) \vert 2\colon p_i,k_i\rangle =
-{e \over {(2\pi)^{3/2}}}\frac{f(k,\rho \colon p_i) }{\sqrt{2k^+}}
\vert 2 \colon p_i, k_i \rangle +
\delta^3(k-k_i)\delta_{\rho{\lambda_i}} \vert 1\colon p_i \rangle\;,
\end{equation}
and
\begin{equation}\label{eigen3}
a^{\dagger}(k,\rho) \vert 1\colon p_i \rangle =
{e \over {(2\pi)^{3/2}}}\frac{f^*(k,\rho \colon p_i) }{\sqrt{2k^+}}
 \vert 1\colon p_i \rangle + \vert 2 \colon p_i, k_i \rangle\;,
\end{equation}
\begin{equation}\label{eigen4}
a^{\dagger}(k,\rho) \vert 2\colon p_i ,k_i\rangle =
{e \over {(2\pi)^{3/2}}}\frac{f^*(k,\rho \colon p_i) }{\sqrt{2k^+}}
\vert 2\colon p_i,k_i \rangle + \vert 3 \colon p_i, k_i ,k_j\rangle\;.
\end{equation}
Coherent states satisfy the following orthonormalization properties:
\begin{align}\label{ortho11}
\langle 1\colon p_f,\sigma_f \vert 1 \colon p_i,\sigma_i\rangle =
\delta^{(3)}(p_i-p_f)\delta_{\sigma_i\sigma_f}
\end{align}
\begin{align}\label{ortho12}
\langle 1 \colon p_f,\sigma_f \vert2\colon p_i,\sigma_i, k_i,\lambda_i\rangle
 ={e \over (2\pi)^{3/2}}\frac{f(k_i,\lambda_i\colon p_i)}{\sqrt{2k_i^+}}
 \delta^{(3)}(p_i-p_f)\delta_{\sigma_i\sigma_f}
\end{align}
\section{Transition matrix element in fock basis for self energy upto $O(e^4)$}
We now calculate $T_3$ which is defined by Eq.~(\ref{T_3}) and corresponds to  Fig. 3. Inserting complete sets of intermediate states in $T_{3}$, we obtain
\begin{align}
T_3(p,p) =T_{3a}+T_{3b}+T_{3c}\nonumber
\end{align}where
\begin{align}\label{T_{3a}}
T_{3a}=&{e^4\over{(2 \pi)^6}}\int{{d^2 {\bf k}_{1\perp}}{d^2{\bf k}_{2\perp}}\over 2p^+ } \int{{dk_1^+}{dk_2^+}\over{32 k_1^+k_2^+p_1^+p_2^+p_3^+}}\nonumber\\&
\frac{\overline u(p,s)[\not\epsilon^{\lambda_1}(k_1)(\not p_1+m)\not\epsilon^{\lambda_2}(k_2)(\not p_2+m)\not\epsilon^{\lambda_2}(k_2)(\not p_1+m)\not\epsilon^{\lambda_1}(k_1)]u(p,s)} {(p^--p_1^--k_1^-)(p^--p_2^--k_1^--k_2^-)(p^--p_1^--k_1^-)}
\end{align}
with
\begin{align}\label{p1}
p_1=p-k_1,\\
\label{p2}
p_2=p-k_1-k_2.
\end{align}
Similarly,
\begin{align}\label{T_{3b}}													
T_{3b}=&{e^4\over{(2 \pi)^6}}\int {{d^2 {\bf k}_{1\perp}}{d^2 {\bf k}_{2\perp}}\over 2p^+ }\int{{dk_1^+}{dk_2^+}\over{32 k_1^+k_2^+p_1^+p_2^+p_3^+}}\nonumber\\&\frac{\overline u(p,s)[\not\epsilon^{\lambda_2}(k_2)(\not p_3+m)\not\epsilon^{\lambda_1}(k_1)(\not p_2+m) \not\epsilon^{\lambda_2}(k_2)(\not p_1+m)\not\epsilon^{\lambda_1}(k_1)] u(p,s)} {(p^--p_1^--k_1^-)(p^--p_3^--k_2^-)(p^--p_2^--k_1^--k_2^-)}
\end{align}
with
\begin{align}\label{p3}
p_3=p-k_2
\end{align}
\begin{align}\label{T_{3c}}
T_{3c}=&\frac{e^4}{(2\pi)^6}\int{{d^2 {\bf k}_{1\perp}}{d^2 {\bf k}_{2\perp}}\over 2p^+ } \int {{dk_1^+}{dk_2^+}\over {32 k_1^+k_2^+p_1^+p_3^+p^+}}\nonumber\\&\frac{\overline u(p,s)[\not\epsilon^{\lambda_2}(k_2)(\not p_3+m) \not\epsilon^{\lambda_2}(k_2) (\not p^\prime+m) \not\epsilon^{\lambda_1}(k_1)(\not p_1+m)\not\epsilon^{\lambda_1}(k_1)]u(p,s)}{(p^--p_1^--k_1^-)(p^--p_2^{\prime -})(p^--p_3^--k_2^-)}
\end{align}
In limit I, Eqs.~(\ref{T_{3a}}) and (\ref{T_{3b}}) can be added such that the denominator reduces to $(p\cdot k_1)^2(p\cdot k_2)$. Using Eqs.~(\ref{uubar}), (\ref{relation1}), (\ref{relation2}) and (\ref{deltam}), we obtain
\begin{align}\label{delta66}
[(\delta m^2)_{3a}+(\delta m^2)_{3b}]^{I}=&-\frac{e^4}{2(2\pi)^6}\int{{d^2 {\bf k}_{1\perp}}{d^2{\bf k}_{2\perp}}}\int {{dk_1^+}{dk_2^+}\over{32 k_1^+k_2^+}}\nonumber\\&\frac{Tr[\not\epsilon^{\lambda_1}(k_1)(\not p+m) \not\epsilon^{\lambda_2}(k_2)(\not p_3+m)\not\epsilon^{\lambda_2}(k_2)(\not p+m)\not\epsilon^{\lambda_1}(k_1)(\not p+m)]}{(p\cdot k_1)^2(p\cdot k_2)}
\end{align}
After calculating the trace, Eq.~(\ref{delta66}) leads to
\begin{align}
& [(\delta m^2)_{3a}+(\delta m^2)_{3b}]^{I}\nonumber\\
&=-{e^4\over{(2\pi)^6}}\int{{d^2{\bf k}_{1\perp}}{d^2{\bf k}_{2\perp}}}\int{{{dk_1^+}\over{k_1^+}}{{dk_2^+}\over {k_2^+}}}\frac{[2(p\cdot \epsilon(k_1))^2(p\cdot \epsilon(k_2))^2-(p\cdot k_2)(p\cdot\epsilon(k_1))^2]}{4(p\cdot k_1)^2(p\cdot k_2)}
\end{align}
Again, in limit II, the denominator in the sum of Eqs.~(\ref{T_{3a}}) and (\ref{T_{3b}}) reduces to $(p\cdot k_1)(p\cdot k_2)^2$ leading to 
\begin{align}\label{delta68}
[(\delta m^2)_{3a}+(\delta m^2)_{3b}]^{II}=&-{e^4\over{2(2 \pi)^6}}\int{{d^2{\bf k}_{1\perp}}{d^2{\bf k}_{2\perp}}}\int {{dk_1^+}{dk_2^+}\over {32 k_1^+k_2^+}}\nonumber\\&\frac{Tr[\not\epsilon^{\lambda_2}(k_2)(\not p+m) \not\epsilon^{\lambda_1}(k_1)(\not p_1+m)\not\epsilon^{\lambda_2}(k_2)(\not p_1+m)\not\epsilon^{\lambda_1}(k_1)(\not p+m)]}{(p\cdot k_1)(p\cdot k_2)^2}
\end{align}
Calculating the trace, Eq.~(\ref{delta68}) reduces to 
\begin{align}
&[(\delta m^2)_{3a}+(\delta m^2)_{3b}]^{II}\nonumber\\
&= -{ e^4\over{(2 \pi)^6}} \int {{d^2{\bf k}_{1\perp}}{d^2{\bf k}_{2\perp}} }\int {{{dk_1^+}\over{k_1^+}}{{dk_2^+}\over {k_2^+}}}\frac{[2(p\cdot \epsilon(k_1))^2(p\cdot \epsilon(k_2))^2-(p\cdot k_1)(p\cdot \epsilon(k_2))^2]}{4(p\cdot k_1)(p\cdot k_2)^2}
\end{align}
In  limit III, we obtain 
\begin{align}
[(\delta m^2)_{3a}+(\delta m^2)_{3b}]^{III}=-{e^4\over{(2 \pi)^6}} \int {{d^2{\bf k}_{1\perp}}{d^2{\bf k}_{2\perp}} }\int {{{dk_1^+}\over{k_1^+}}{{dk_2^+}\over {k_2^+}}}\frac{(p\cdot \epsilon(k_1))^2(p\cdot \epsilon(k_2))^2}{2(p\cdot k_1)^2(p\cdot k_2)}
\end{align}
Similarly, $T_{3c}$ leads to 
\begin{align}\label{delta3c}
\delta m^2_{3c}=&\frac{e^4}{2(2\pi)^6}\int {{d^2 {\bf k}_{1\perp}}{d^2 {\bf k}_{2\perp}}}\int {{dk_1^+}{dk_2^+}\over {32k_1^+k_2^+p_1^+p_3^+p^+}}\nonumber\\ &\frac{Tr[\not\epsilon^{\lambda_2}(k_2)(\not p_3+m) \not\epsilon^{\lambda_2}(k_2) (\not p^\prime+m)\not\epsilon^{\lambda_1}(k_1)(\not p_1+m)\not\epsilon^{\lambda_1}(k_1)(\not p+m)]} {(p^--p_1^--k_1^-)(p^--p_2^{\prime -})(p^--p_3^--k_2^-)}
\end{align}
where $p_1$ and $p_3$ are defined by Eqs.~(\ref{p1}) and (\ref{p3}) and $p_2^\prime=p$. Note that  this  diagram is one-particle reducible, and therefore the energy denominator associated with the single-particle state vanishes. We shall use the Heitler method \cite{HEIT54} for evaluating all such integrals. Using this method, we write \cite{MUS91,HEIT54} 
\begin{align}\label{Heitler1}
D=&\frac{1}{(p^--p_2^{\prime-})(p^--p_1^--k_1^-)(p^--p_3^--k_2^-)}\nonumber\\
=&\int dp^{\prime-}\delta(p^{\prime-}-p^-) \frac{\mathcal P} {(p^{\prime-}-p_2^{\prime-})(p^{\prime-}-p_1^--k_1^-)(p^{\prime-}-p_3^--k_2^-)},
\end{align}
Using the relation between distributions
\begin{align}
\frac{\mathcal P}{(p^{\prime-}-p_2^{\prime-})}\delta(p^{\prime-}-p^-)=-{1\over2} \delta^\prime(p^{\prime-}-p^-).
\end{align}
and integrating by parts we obtain
\begin{align}\label{Heitler2}
D=&{1\over2}\int dp^{\prime-}\delta(p^{\prime-}-p^-){d\over dp^{\prime-}}\biggl[{1\over{(p^{\prime-}-p_1^--k_1^-)(p^{\prime-}-p_3^--k_2^-)}}\biggr]
\nonumber\\
=&-\frac{1}{2(p^--p_1^--k_1^-)^2(p^--p_3^--k_2^-)}-\frac{1}{2(p^--p_1^--k_1^-)(p^--p_3^--k_2^-)^2}
\end{align}
Using Eq.~(\ref{relation1}) in Eq.~(\ref{Heitler2}) we obtain
\begin{align}
D=\frac{p_1^+}{2(p\cdot k_1)^2(p\cdot k_2)}+\frac{p_3^+}{2(p\cdot k_1)(p\cdot k_2)^2}
\end{align}
Thus, in limit I, Eq.~(\ref{delta3c}) becomes
\begin{align}
[(\delta m^2)_{3c}]^{I}=&{e^4\over{(2 \pi)^6}} \int {{d^2{\bf k}_{1\perp}}{d^2{\bf k}_{2\perp}} }\int {{{dk_1^+}\over{k_1^+}}{{dk_2^+}\over {k_2^+}}}\frac{2(p\cdot \epsilon(k_1))^2(p\cdot \epsilon(k_2))^2-(p\cdot k_2)(p\cdot \epsilon(k_1))^2}{8p^+}\nonumber\\&\biggl[\frac{p^+}{(p\cdot k_1)^2(p\cdot k_2)}+\frac{p_3^+}{(p\cdot k_1)(p\cdot k_2)^2}\biggr]
\end{align}
Similarly,  in  limit II,  Eq.~(\ref{delta3c}) leads to
\begin{align}
[(\delta m^2)_{3c}]^{II}=&\frac{e^4}{(2\pi)^6}\int{{d^2{\bf k}_{1\perp}}{d^2{\bf k}_{2\perp}}}\int{{{dk_1^+}\over {k_1^+}}{{dk_2^+}\over{k_2^+}}}\frac{2(p\cdot \epsilon(k_1))^2(p\cdot \epsilon(k_2))^2-(p\cdot k_1)(p\cdot \epsilon(k_2))^2}{8p^+}\nonumber\\&\biggl[\frac{p_1^+}{(p\cdot k_1)^2(p\cdot k_2)}+\frac{p^+}{(p\cdot k_1)(p\cdot k_2)^2}\biggr]
\end{align}
and in limit  III, it gives
\begin{align}
[(\delta m^2)_{3c}]^{III}=&\frac{e^4}{(2\pi)^6}\int{{d^2{\bf k}_{1\perp}}{d^2{\bf k}_{2\perp}}}\int{{{dk_1^+} \over{k_1^+}}{{dk_2^+}\over{k_2^+}}}\nonumber\\ &\biggl[\frac{(p\cdot \epsilon(k_1))^2(p\cdot \epsilon(k_2))^2}{4(p\cdot k_1)^2 (p\cdot k_2)}+\frac{(p\cdot \epsilon(k_1))^2(p\cdot \epsilon(k_2))^2}{4(p\cdot k_1)(p\cdot k_2)^2}\biggr]
\end{align}
The traces are calculated using Mathematica. 

The contribution corresponding diagrams in Fig. 4 is given by 
\begin{displaymath}
(\delta m^2)_4=(\delta m^2)_{4a}+(\delta m^2)_{4b}+(\delta m^2)_{4c}
\end{displaymath}
In limit I, $(\delta m^2)_{4a}$ reduces to
\begin{align}\label{4a1}
[(\delta m^2)_{4a}]^{I}=&-{e^4\over{(2 \pi)^6}}\int{{d^2{\bf k}_{1\perp}}{d^2{\bf k}_{2\perp}}}\int{{{dk_1^+}\over{k_1^+}}{{dk_2^+} \over{k_2^+}}}\nonumber\\ &\frac{[2p^+(p\cdot \epsilon(k_1))(p\cdot \epsilon(k_2))(\epsilon(k_1)\cdot\epsilon(k_2))+k_2^+(p\cdot \epsilon(k_1))^2]}{8p^+(p\cdot k_1)(p\cdot k_2)}
\end{align}
Note that $(\delta m^2)_{4a}$ is not IR divergent when $p\cdot k_1 \neq 0$ even if $p\cdot k_2=0$. \\ 
In limit II, $(\delta m^2)_{4b}$ reduces to
\begin{align}\label{4b2}
[(\delta m^2)_{4b}]^{II}=&\frac{e^4}{(2\pi)^6}\int{{d^2{\bf k}_{1\perp}}{d^2{\bf k}_{2\perp}}}\int{{{dk_1^+} \over{k_1^+}}{{dk_2^+} \over{k_2^+}}}\nonumber\\ &\frac{[2p_1^+(p\cdot \epsilon(k_1))(p\cdot \epsilon(k_2))(\epsilon(k_1)\cdot\epsilon(k_2))-k_1^+(p\cdot \epsilon(k_2))^2]}{8p_1^+(p\cdot k_1)(p\cdot k_2)}
\end{align}
Note that $(\delta m^2)_{4b}$ is not IR divergent when $p\cdot k_2 \neq 0$ even if $p\cdot k_1=0$. \\
In limit III and after adding the contributions from Figs. 4(a) and 4(b), we get
\begin{align}\label{4a3}
[(\delta m^2)_{4a}+(\delta m^2)_{4b}]^{III}=-{e^4\over{(2\pi)^6}}\int{{d^2{\bf k}_{1\perp}}{d^2{\bf k}_{2\perp}} }\int{{{dk_1^+}\over {k_1^+}}{{dk_2^+}\over{k_2^+}}}\frac{[(p\cdot \epsilon(k_1))(p\cdot \epsilon(k_2)) (\epsilon(k_1)\cdot\epsilon(k_2))]}{4(p\cdot k_1)(p\cdot k_2)}
\end{align}
For $(\delta m^2)_{4c}$, we use the Heitler method illustrated in Eqs.~(\ref{Heitler1})-(\ref{Heitler2}) and obtain 
\begin{align}\label{4c}
(\delta m^2)_{4c}=&\frac{e^4}{2(2\pi)^6}\int{{d^2 {\bf k}_{1\perp}}{d^2 {\bf k}_{2\perp}}}\int{{dk_1^+}{dk_2^+}\over {32 k_1^+k_2^+p^+}}\nonumber\\ &\frac{Tr[\not\epsilon^{\lambda_1}(k_1)(\not p+m)\not\epsilon^{\lambda_1}(k_1)(\not p+m) \not\epsilon^{\lambda_2}(k_2)\gamma^+\not\epsilon^{\lambda_2}(k_2)(\not p+m)]}{(p\cdot k_1)^2}
\end{align}
which finally leads to 
\begin{align}\label{delta4cir}
[(\delta m^2)_{4c}]^{I}=-{e^4\over{(2\pi)^6}}\int{{d^2{\bf k}_{1\perp}}{d^2{\bf k}_{2\perp}}}\int{{{dk_1^+}\over {k_1^+}}{{dk_2^+}\over {k_2^+}}}\frac{{(p\cdot \epsilon(k_1))}^2}{8(p\cdot k_1)^2}
\end{align}

Similarly, contribution from Figs. 5(a)-(c) is given by 
\begin{align}\label{delta5}
(\delta m^2)_5=(\delta m^2)_{5a}+(\delta m^2)_{5b}+(\delta m^2)_{5c}\nonumber
\end{align}
In limit I, $(\delta m^2)_{5a}$ reduces to 
\begin{align}
[(\delta m^2)_{5a}]^{I}=&-{e^4\over{(2 \pi)^6}} \int {{d^2{\bf k}_{1\perp}}{d^2{\bf k}_{2\perp}}}\int{{{dk_1^+} \over{k_1^+}}{{dk_2^+}\over{k_2^+}}}\nonumber\\&\frac{[2p^+(p\cdot \epsilon(k_1))(p\cdot \epsilon(k_2)) (\epsilon(k_1)\cdot\epsilon(k_2))+k_2^+(p\cdot \epsilon(k_1))^2]}{8p^+(p\cdot k_1)(p\cdot k_2)}
\end{align}
Similarly, in limit II we get
\begin{align}
[(\delta m^2)_{5a}]^{II}=&-{e^4\over{(2\pi)^6}}\int {{d^2{\bf k}_{1\perp}}{d^2{\bf k}_{2\perp}}}\int{{{dk_1^+} \over{k_1^+}} {{dk_2^+}\over{k_2^+}}}\nonumber\\&\frac{[2p^+(p\cdot \epsilon(k_1))(p\cdot \epsilon(k_2))(\epsilon(k_1)\cdot\epsilon(k_2)) +k_1^+(p\cdot \epsilon(k_2))^2]}{8p_1^+(p\cdot k_1)(p\cdot k_2)}
\end{align}
and in limit III we get
\begin{align}\label{5a3}
[(\delta m^2)_{5a}]^{III}={e^4\over {(2 \pi)^6}} \int {{d^2{\bf k}_{1\perp}}{d^2{\bf k}_{2\perp}} }\int{{{dk_1^+}\over {k_1^+}}{{dk_2^+}\over {k_2^+}}}\frac{[(p\cdot \epsilon(k_1))(p\cdot \epsilon(k_2))(\epsilon(k_1)\cdot\epsilon(k_2))]}{4(p\cdot k_1)(p\cdot k_2)}
\end{align}
Taking limit I $(\delta m^2)_{5b}$ reduces to
\begin{align}
[(\delta m^2)_{5b}]^{I}=&-{e^4\over{(2 \pi)^6}} \int {{d^2{\bf k}_{1\perp}}{d^2{\bf k}_{2\perp}}}\int{{{dk_1^+} \over{k_1^+}}{{dk_2^+}\over{k_2^+}}}\nonumber\\&\frac{[2p_3^+(p\cdot \epsilon(k_1))(p\cdot \epsilon(k_2)) (\epsilon(k_1)\cdot\epsilon(k_2))-k_2^+(p\cdot \epsilon(k_1))^2]}{8p_3^+(p\cdot k_1)(p\cdot k_2)}
\end{align}
Taking limit II we get
\begin{align}
[(\delta m^2)_{5b}]^{II}=&-{e^4\over{(2\pi)^6}} \int {{d^2{\bf k}_{1\perp}}{d^2{\bf k}_{2\perp}}}\int{{{dk_1^+} \over{k_1^+}}{{dk_2^+}\over{k_2^+}}}\nonumber\\&\frac{[2p_1^+(p\cdot \epsilon(k_1))(p\cdot \epsilon(k_2)) (\epsilon(k_1)\cdot\epsilon(k_2))-k_1^+(p\cdot \epsilon(k_2))^2]}{8p_1^+(p\cdot k_1)(p\cdot k_2)}
\end{align}
and taking limit III we get
\begin{align}\label{5b3}
[(\delta m^2)_{5b}]^{III}=-{e^4\over {(2 \pi)^6}} \int {{d^2{\bf k}_{1\perp}}{d^2{\bf k}_{2\perp}} }\int{{{dk_1^+}\over {k_1^+}}{{dk_2^+}\over {k_2^+}}}\frac{[(p\cdot \epsilon(k_1))(p\cdot \epsilon(k_2))(\epsilon(k_1)\cdot\epsilon(k_2))]}{4(p\cdot k_1)(p\cdot k_2)}
\end{align}
$(\delta m^2)_{5c}$ in the limit I, reduces to
\begin{align}\label{5c1}
[(\delta m^2)_{5c}]^{I}= {e^4\over{(2 \pi)^6}} \int {{d^2{\bf k}_{1\perp}}{d^2{\bf k}_{2\perp}} }\int {{{dk_1^+}\over {k_1^+}}{{dk_2^+}\over {k_2^+}}}\frac{{(p\cdot \epsilon(k_1))}^2}{{4(p\cdot k_1)}^2}
\end{align}

Contribution of Figs. 6(a)-(c) is given by
\begin{align}
(\delta m^2)_{6}=(\delta m^2)_{6a}+(\delta m^2)_{6b}+(\delta m^2)_{6c}\nonumber
\end{align}
In limit I, $(\delta m^2)_{6a}$ reduces to
\begin{align}
[(\delta m^2)_{6a}]^{I}=&-{e^4\over{(2 \pi)^6}}\int{{d^2{\bf k}_{1\perp}}{d^2{\bf k}_{2\perp}}} \int{{{dk_1^+} \over{k_1^+}}{{dk_2^+}\over{k_2^+}}}\nonumber\\&\frac{[2p^+(p\cdot \epsilon(k_1))(p\cdot\epsilon(k_2))(\epsilon(k_1) \cdot\epsilon(k_2)) +k_2^+(p\cdot\epsilon(k_1))^2]}{8p^+(p\cdot k_1)(p\cdot k_2)}
\end{align}
Note that $(\delta m^2)_{6a}$ is not IR divergent when $p\cdot k_1 \neq 0$ even if $p\cdot k_2=0$. \\
In limit II, $(\delta m^2)_{6b}$ reduces to
\begin{align}
[(\delta m^2)_{6b}]^{II}=&-{e^4\over{(2\pi)^6}}\int{{d^2{\bf k}_{1\perp}}{d^2{\bf k}_{2\perp}}}\int{{{dk_1^+} \over{k_1^+}}{{dk_2^+}\over{k_2^+}}}\nonumber\\&\frac{[2p_1^+(p\cdot\epsilon(k_1))(p\cdot\epsilon(k_2))(\epsilon(k_1)\cdot\epsilon(k_2))-k_1^+(p\cdot\epsilon(k_2))^2]}{8p_1^+(p\cdot k_1)(p\cdot k_2)}
\end{align}
Note that $(\delta m^2)_{6b}$ is not IR divergent when $p\cdot k_2 \neq 0$ even if $p\cdot k_1=0$. \\
In limit III, sum of $(\delta m^2)_{6a}$ and $(\delta m^2)_{6b}$ reduces to
\begin{align}\label{6a3}
[(\delta m^2)_{6a}+(\delta m^2)_{6b}]^{III}=-{e^4\over{(2\pi)^6}}\int{{d^2{\bf k}_{1\perp}}{d^2{\bf k}_{2\perp}}} \int{{{dk_1^+}\over{k_1^+}}{{dk_2^+}\over{k_2^+}}}\frac{[(p\cdot\epsilon(k_1))(p\cdot\epsilon(k_2)) (\epsilon(k_1)\cdot\epsilon(k_2))]}{4(p\cdot k_1)(p\cdot k_2)}
\end{align}
For Fig. 6(c) we use Heitler method \cite{MUS91} to obtain the following result in limit I
\begin{align}
[(\delta m^2)_{6c}]^I=-{e^4\over{(2 \pi)^6}}\int{{d^2{\bf k}_{1\perp}}{d^2{\bf k}_{2\perp}}}\int {{{dk_1^+}\over{k_1^+}}{{dk_2^+}\over{k_2^+}}}\frac{{(p\cdot\epsilon(k_1))}^2}{8(p\cdot k_1)^2}
\end{align}
\vskip 0.5cm
\section{Transition matrix in coherent state basis}
We will now present the calculation of ${(\delta m^2)^{\prime(2)}}$ where prime denotes the extra contribution arising due to use of coherent state basis.
Contribution corresponding to Fig. 8 can be written as
\begin{align}
(\delta m^2)_{8}^\prime=(\delta m^2)_{8a}^\prime+(\delta m^2)_{8b}^\prime+(\delta m^2)_{8c}^\prime+(\delta m^2)_{8d}^\prime+(\delta m^2)_{8e}^\prime+(\delta m^2)_{8f}^\prime\nonumber
\end{align}
where
\begin{align}\label{delta8a}
{(\delta m^2)}_{8a}^\prime=&\frac{e^4}{2(2\pi)^6}\int{{d^2{\bf k}_{1\perp}}{d^2{\bf k}_{2\perp}}}\int{{dk_1^+}{dk_2^+} \over {16k_1^+k_2^+}}\nonumber\\ &\frac{Tr[\not\epsilon^{\lambda_2}(k_2)(\not p_3+m)\not\epsilon^{\lambda_1}(k_1)(\not p_1+m)\not\epsilon^{\lambda_1}(k_1)(\not p+m)] (p\cdot\epsilon^{\lambda_1}(k_1))\Theta_\Delta(k_1)}{(p\cdot k_1)^2[(p\cdot k_1)+(p\cdot k_2)-(k_1\cdot k_2)]}
\end{align}
\begin{align}\label{delta8b}
(\delta m^2)_{8b}^\prime=&\frac{e^4}{2(2\pi)^6}\int{{d^2{\bf k}_{1\perp}}{d^2{\bf k}_{2\perp}}}\int{{dk_1^+}{dk_2^+} \over {16k_1^+k_2^+}}\nonumber\\&\frac{Tr[\not\epsilon^{\lambda_2}(k_2)(\not p_3+m)\not\epsilon^{\lambda_1}(k_1)(\not p_3+m)\not\epsilon^{\lambda_2}(k_2)(\not p+m)](p\cdot\epsilon^{\lambda_1}(k_1))\Theta_\Delta (k_1)}{(p\cdot k_1)(p\cdot k_2)[(p\cdot k_1)+(p\cdot k_2)-(k_1\cdot k_2)]}
\end{align}
Calculating the trace, adding Eqs.~(\ref{delta8a}) and (\ref{delta8b}) and taking limit I, we obtain  Eqs.~(\ref{coh8a}). Similarly on can obtain the expressions for Figs. 8(c)-8(f) in appropriate limits. 
Fig. 9 corresponds to the transition matrix element $T_9^\prime$ and its contribution is given by
\begin{align}
(\delta m^2)_{9}^\prime=(\delta m^2)_{9a}^\prime+(\delta m^2)_{9b}^\prime+(\delta m^2)_{9c}^\prime+(\delta m^2)_{9d}^\prime\nonumber
\end{align}
where
\begin{align}\label{coh9a}
[(\delta m^2)_{9a}^\prime]^{III}=-{e^4\over{(2 \pi)^6}}\int {{d^2{\bf k}_{1\perp}}{d^2{\bf k}_{2\perp}}}\int{{{dk_1^+} \over{k_1^+}}{{dk_2^+}\over {k_2^+}}}\frac{(p\cdot\epsilon(k_1))^2(p\cdot\epsilon(k_2))^2}{4(p\cdot k_1)^2(p\cdot k_2)} \Theta_{\Delta}(k_1)\Theta_{\Delta}(k_2)
\end{align}
\begin{align}\label{coh9b}
[(\delta m^2)_{9b}^\prime]^{III}=-{e^4\over{(2 \pi)^6}}\int {{d^2{\bf k}_{1\perp}}{d^2{\bf k}_{2\perp}}}\int{{{dk_1^+} \over{k_1^+}}{{dk_2^+}\over {k_2^+}}}\frac{(p\cdot\epsilon(k_1))^2(p\cdot\epsilon(k_2))^2}{4(p\cdot k_1)^2(p\cdot k_2)} \Theta_{\Delta}(k_1)\Theta_{\Delta}(k_2)
\end{align}
\begin{align}\label{coh9c}
[(\delta m^2)_{9c}^\prime+{(\delta m^2)}_{9d}^\prime]^{III}=&{e^4\over{(2 \pi)^6}}\int{{d^2{\bf k}_{1\perp}}{d^2{\bf k}_{2\perp}}}\int{{{dk_1^+} \over{k_1^+}}{{dk_2^+}\over{k_2^+}}}\biggl[\frac{(p\cdot \epsilon(k_1))^2(p\cdot \epsilon(k_2))^2}{4(p\cdot k_1)^2(p\cdot k_2)}\nonumber\\&+\frac{(p\cdot \epsilon(k_1))^2(p\cdot \epsilon(k_2))^2}{4(p\cdot k_1)(p\cdot k_2)^2}\biggr]\Theta_{\Delta}(k_1)\Theta_{\Delta}(k_2)
\end{align}
Here we have used Heitler method to get the above result.

Similarly, contribution coming from diagrams in Figs. 10 and 11 can be easily evaluated by putting the appropriate limits. 

\end{document}